\documentclass[aps,prx,twocolumn,superscriptaddress,preprintnumbers]{revtex4-2}

\usepackage[T1]{fontenc}
\usepackage{amssymb}
\usepackage{graphicx}
\usepackage{color}
\usepackage{caption}
\usepackage{hyperref}
\usepackage{amsthm}
\usepackage{amsmath}
\usepackage{braket}
\usepackage{appendix}
\usepackage{subcaption}
\usepackage{comment}
\usepackage{bbold}
\usepackage{multirow}
\usepackage{booktabs}

\usepackage{tikz}
\usetikzlibrary{patterns}
\usetikzlibrary{arrows.meta,
                positioning}

\usepackage{physics}

\usepackage{amsthm}
\usepackage{enumitem}
\usepackage{comment}

\begin{document}


\title{
Interplay between local and non-local frustration in the 1D ANNNI chain I \\ The even case
}

\author{G. Torre}
\affiliation{Institut Ru\dj er Bo\v{s}kovi\'c, Bijeni\v{c}ka cesta 54, 10000 Zagreb, Croatia}

\author{A. G. Catalano}
\affiliation{Institut Ru\dj er Bo\v{s}kovi\'c, Bijeni\v{c}ka cesta 54, 10000 Zagreb, Croatia}
\affiliation{Universit\'e de Strasbourg, 4 Rue Blaise Pascal, 67081 Strasbourg, France}

\author{S. B. Kožić}
\affiliation{Institut Ru\dj er Bo\v{s}kovi\'c, Bijeni\v{c}ka cesta 54, 10000 Zagreb, Croatia}

\author{F. Franchini}
\email{fabio@irb.hr}
\affiliation{Institut Ru\dj er Bo\v{s}kovi\'c, Bijeni\v{c}ka cesta 54, 10000 Zagreb, Croatia}

\author{S. M. Giampaolo}
\email{sgiampa@irb.hr}
\affiliation{Institut Ru\dj er Bo\v{s}kovi\'c, Bijeni\v{c}ka cesta 54, 10000 Zagreb, Croatia}

\preprint{RBI-ThPhys-2024-11}


\begin{abstract}
We consider the effects of the competition between different sources of frustration in 1D spin chains through the analysis of the paradigmatic  ANNNI model, which possesses an extensive amount of frustration of local origin due to the competition between nearest and next-to-nearest neighbor interactions. An additional, non-extensive amount of topological frustration can be added by applying suitable boundary conditions and we show that this seemingly subdominant contribution significantly affects the model. Choosing periodic boundary conditions with an {\it even} number of sites not divisible by $4$ and using the entanglement entropy as a probe, we demonstrate that in one of the model's phases the ground state can be characterized as hosting two (almost) independent excitations. Thus, not only we show an intriguing interplay between different types of frustration, but also manage to propose a non-trivial quasi-particle interpretation for it.

\end{abstract}

\maketitle

\section{Introduction}

The field of quantum complex systems emerged as the natural extension of classical statistical physics.
Therefore, it is not surprising that the first studies were focused on the analysis of quantities such as correlation functions and local order parameters that, accordingly with Landau's theory~\cite{Landau1937}, characterize the different phases of a physical system.
This approach achieved remarkable successes, but the inherent non-locality of quantum mechanics gave rise to a much broader phenomenology than that of classical systems, which was impossible to fully encompass within this framework.
A notable example of this fact is represented by topologically ordered phases that are characterized by robust ground state degeneracies~\cite{Wen&Niu1990} and can be unveiled by topological invariants that remain unaffected by continuous deformations of the Hamiltonian parameters~\cite{Wen1990, Zeng2015, Stanescu2016, Moessner2021}. 
In these phases, the global structures of the ground states are connected to non-local correlations whose presence can be highlighted by the entanglement properties~\cite{Wen2002, Hamma2005, Chen2010, Wen2013, Wen2017, Wen2019}. 

However, such order represents just one example, even if the most widely known, of the different phenomenologies lying outside Landau's theory which are continually unveiled.
Another one, which represents the topic of the present work, is the so-called {\it topological frustration} (TF).
The concept of frustration in quantum mechanics must be handled with care.
By definition, a many-body system is frustrated when it is impossible to minimize simultaneously all its local energy constrains. 
Due to the non-commuting nature of quantum mechanics, with few exceptions such as the frustration free models~\cite{Shor2016, Tong2021} or systems at a factorization point~\cite{fact1, fact2, fact3}, almost all quantum systems include some amount of frustration~\cite{Wolf2003, Giampaolo2011, Giampaolo2013, Giampaolo2015}. 
But, in line with recent usage in the scientific community, in the following we restrict the concept of frustration to effects induced by competing interactions and/or geometrical properties, extending it directly from classical systems~\cite{Toulouse1977, Vannimenus1977}.

Returning to TF, initially it has been examined in models featuring antiferromagnetic (AFM) nearest neighbor coupling and implemented on lattices with frustrated boundary conditions (FBCs)~\cite{Dong2016}. 
FBCs refer to the coexistence of periodic boundary conditions (PBCs) and structures comprised of an odd number of spins.
In the classical limit, where all 
terms in the Hamiltonian commutes each other, these systems develop an extensive degeneracy of the ground state which is typically lifted away when quantum terms are taken into account~\cite{Sen2008, Maric2022_fate, Catalano2022}.
As a result, a gapless band structure replaces the gapped phase featured by the same systems in the same region of parameter spaces but with different boundary conditions.
FBC induce peculiar phenomenologies, revealed, for instance, by an increment in complexity of the ground state~\cite{Odavic2023}, the suppression of the usual AFM order~\cite{Maric2020_destroy, Maric2020_neworder}, a huge sensitivity of the dynamical properties to small local perturbation~\cite{Torre2022}, etc.
As for the topologically ordered phase, also the topological nature of such phenomenology can be unveiled by the analysis of the entanglement entropy~\cite{Giampaolo2019}, which clearly shows a topological contribution~\cite{Torre2023}.
All these results emerged in the past few years and have been quite unexpected, challenging the prevailing belief that boundary conditions could not produce effects lasting up to the thermodynamic limit.

However, the analysis carried out so far focused on one-dimensional spin chains with just short-range (nearest-neighbor) interactions. 
These systems, by their nature, could not show any source of local frustration, i.e. generated by different interactions competing on the same sites.
Therefore, TF has been the only type of frustration and, being associated with boundary conditions, provides a sub-extensive amount of frustration.
With this work we start extending the analysis to more complex systems, those that present also local sources of frustration due to the interplay between neighboring and non-neighboring spin pairs interactions, and investigate whether TF can influence these systems as well. We consider the ANNNI model, being the simplest one in this respect. It was initially introduced as a theoretical model to understand numerous experimental observations~\cite{Selke88,Suzuki2013}, and it is now regaining attention as the right playground to test machine learning algorithms, due to its rich phase diagram and the lack of a general analytical solution~\cite{Monaco2023, Canabarro2019}. Furthermore, it is amenable to experimental implementation with segmented ion traps~\cite{Zippilli2014}. The Hamiltonian of the model for $L$ spins reads
\begin{equation}
    \label{eq:Hamiltonian}
    H=J_1\sum_{i=1}^L \sigma_{i}^x\sigma_{i+1}^x + J_2\sum_{i=1}^L \sigma_{i}^x\sigma_{i+2}^x +  h\sum_{i=1}^L\sigma_i^z,
\end{equation}
where $\sigma_{i}^\alpha$, with $\alpha= x,\,y,\,z$ are the Pauli operators acting on the $i$-th spin, $J_1$ and $J_2$ are respectively the nearest and the next-to-nearest neighbor interactions, and $h$ is a transverse magnetic field. 
Unless stated otherwise, we assume periodic boundary conditions, i.e. $\sigma_{i}^\alpha\equiv\sigma_{i+L}^\alpha$.

\begin{figure}
	\includegraphics[width=0.99\columnwidth]{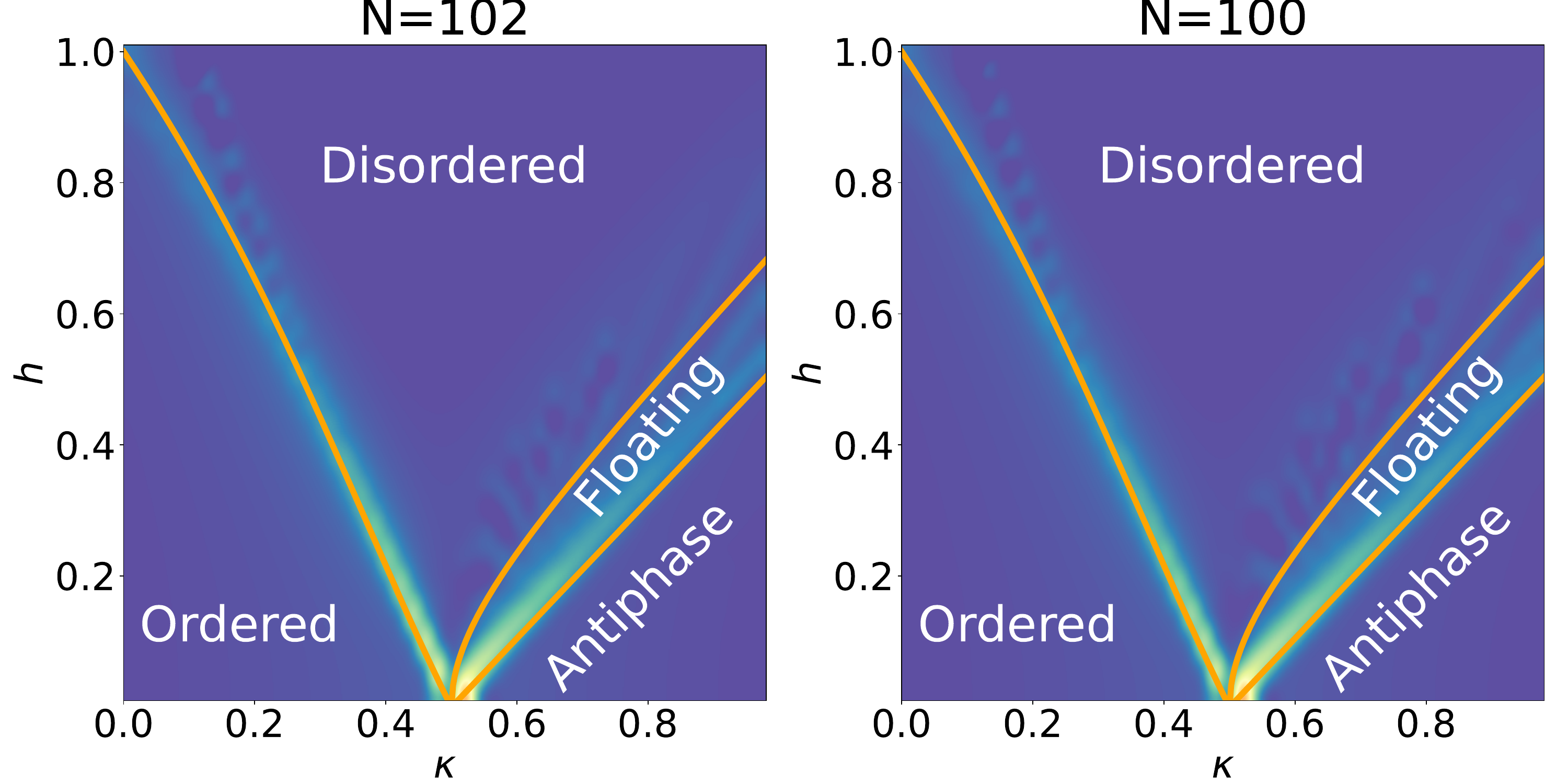}
	\caption{Phase diagram of the ANNNI model in the $(h/J_1,\kappa)$ plane in absence ($L=100$, left panel) and in presence ($L=102$, right panel) of topological frustration, obtained by analyzing the second derivative of the ground-state energy with respect to $h$.
		The solid orange lines represent the quantum phase transitions detected by discontinuities in the energy derivative.
		While the phase diagram, originally drawn in \cite{Suzuki2013}, is unaffected by the different boundary conditions considered, we will show that the antiphase has different properties for $L=4N$ and $L=4N+2$.}
	\label{fig:PhaseDiagram}
\end{figure}

Indeed, an AFM next-to-nearest neighbors coupling ($J_2>0$) induces an extensive frustration of local origin, regardless of the choice of the boundary conditions and of the sign of $J_1$ and the different interaction terms in eq.~\eqref{eq:Hamiltonian} result into a rich phase diagram (see Fig.~\ref{fig:PhaseDiagram}), with four different phases \cite{Suzuki2013,Canabarro2019}, which we will present in the next section.

In this work we add TF, and we will show that this intensive source of frustration 
affect the model, in particular in the so-called {\it antiphase} where $J_2$ is the dominant interaction. 
To prove such a result, we focus on the bipartite entanglement entropy (EE). 
In \cite{Giampaolo2019, Odavic2023} it was shown that in the Ising chain (eq.~\eqref{eq:Hamiltonian} with $J_2=0$) TF, induced by FBC, adds an amount of EE corresponding to the presence of a single delocalized excitation 
With a finite $J_2>0$ we will prove that TF can be induced by choosing a chain with an even number of sites non-divisible by $4$ and this results in a contribution to the EE compatible with that of two delocalized excitations, obeying a Pauli principle that prevents them from occupying the same momentum state. 
The case with an odd number of site will be considered in the next article in this series.

After discussing the phase diagram of the ANNNI model in Sec.~\ref{sec:phasediagram}, we proceed with our analysis of the EE.
In Section~\ref{pert_theory}, with details provided in Appendix~\ref{sec:app_analytical}, we derive an analytical expression for the system's unique ground state close to the classical line $h=0$, by exploiting graph theory within a perturbative approach.
This result enables us to provide an analytical formula for the EE in the thermodynamic limit, which can be cast as a sum of two contributions: one coming from the double $\mathcal{Z}_2$ symmetry of flipping each spin on each even/odd sub-lattice, and the other  revealing the presence of two delocalized excitations within the system's ground state.
We obtained such an expression by exploiting the fact that for large systems the elements of the reduced density matrix can be recasted in terms of Chebyshev polynomials and thanks to this formulation he asymptotic eigenvalues can be evaluated with the use of the so-called Cayley-Hamilton theorem~\cite{Gantmacher}.

Subsequently, we will move beyond the perturbative regime and consider the whole phase diagram of the model using a DMRG numerical approach.
In Section~\ref{num_analys} we will present the value of the EE that we extracted for different system sizes and Hamiltonian parameters and demonstrate that, in the thermodynamic limit, it can be written as the sum of two contributions: one that coincides with the amount of bipartite entanglement without TF, and one directly associated with the presence of TF which matches the analytical expression we obtained analytically close to the classical line.
This result is consistent with the same separation obtained for TF induced by FBCs for the EE in~\cite{Giampaolo2019} and for the non-stabilizerness entropy in~\cite{Odavic2023}. 
Hence, it substantiates the conjecture that every quantum resource of a TF system, in the thermodynamic limit, can be decomposed as the sum of the value of the resource for the corresponding unfrustrated model and the topological contribution measured in proximity to the classical point.

At the end, in Section~\ref{conclusion} we discuss the results and draw our conclusions, while all the technical details are provided in the Appendices.

\section{The Phase Diagram of the ANNNI Model}
\label{sec:phasediagram}

In Fig.~\ref{fig:PhaseDiagram} we draw the phase diagram of the ANNNI chain, that we obtain by the analysis of the energy discontinuities.

Without the external transverse field ($h=0$) every term of the Hamiltonian in eq.~\eqref{eq:Hamiltonian} mutually commute and the model can be considered as classical. 
Introduce the dimensionless coupling $\kappa \equiv J_2/\vert J_1\vert$, in such a limit, we can identify two phases separated by a multicritical point located at $\kappa = \kappa_c = 1/2$.
For $\kappa < \kappa_c$ we have a standard ferromagnetic ($J_1<0$) or AFM ($J_1>0$) ordered phase, while for $\kappa > k_c$ the system endeavors to arrange itself so that each spin has one neighbor spin aligned and one anti-aligned (see also Section~\ref{pert_theory}). This order is referred to as the antiphase~\cite{Suzuki2013}. 

Turning back on the magnetic field, three line of phase transition originate from the multicritical point $(h,\kappa)=(0,1/2)$. One line extends up to $(|h|,\kappa)=(\vert J_1 \vert,0)$, and is a phase transition of the Ising-type between an ordered and a paramagnetic disordered phase~\cite{Beccaria06}. To the right of this line, increasing 
$\kappa$ for $h\neq0$ we encounter two more transition lines. The first marks a Berezinskii–Kosterlitz–Thouless (BKT) transition between the disordered and a floating phase~\cite{Beccaria07, Bak1982}. The latter is a gapless phase described by a Luttinger liquid with algebraic incommensurate correlations~\cite{Giamarchi}. The second line marks a commensurate-to-incommensurate transition between the floating and the antiphase.
In this latter region the AFM next-to-nearest interaction $J_2$ is the dominant one, and we will focus on this antiphase to study the interplay between different kinds of frustration. We will show that, in this phase TF can be induced not only when the system is made of an odd number of spins, but also when the length of the chain is equal to $L=4 N+2$, for some $N\in \mathbb{N}$. 
For the sake of clarity, in this paper we will focus on this even site configuration, leaving the investigation of the case with an odd number of sites to the next work.

\begin{figure}
	\includegraphics[width=1.0\columnwidth]{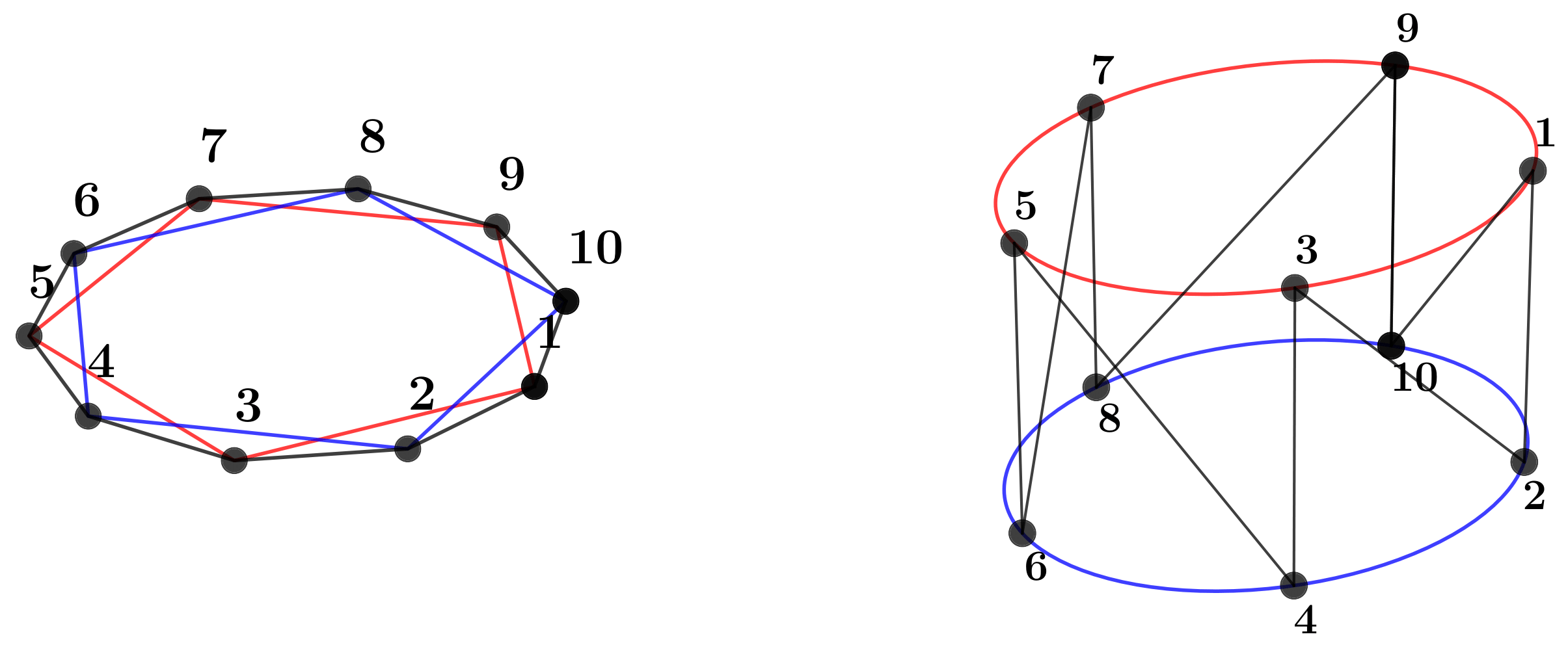}
	\caption{ANNNI chain for $L=10$ lattice sites viewed as two interacting topologically frustrated Ising chains of length $L' = L/2$, composed of the odd and even spins, respectively. When $J_1 = 0$, the two rings are disconnected.}
	\label{fig:ANNNI_chain_decomposition}
\end{figure}

\section{Analytical results close to the classical point}\label{pert_theory}

Among other properties, the phase transition across the multicritical point $\kappa_c=0.5$ between the ordered phase and the antiphase at $h=0$ manifests itself through a change in the dimensionality of the ground state manifold (GSM).
For $\kappa < 1/2$, the system shows a two-fold degenerate GSM for any even $L$, generated by the two orthogonal N\'eel states. Instead, in the antiphase ($\kappa > 1/2$) two different situations arise, depending on the chain lengths.
When $L=4N$, the system is characterized by a four-fold GSM, composed by periodic configurations with 4-sites periodicity ($\ket{\uparrow\uparrow\downarrow\downarrow\uparrow\uparrow\downarrow\downarrow\dots}$).
On the contrary, for $L=4N+2$, the GSM shows a massive ground state degeneracy in which the number of elements scales quadratically with the system size. 
To understand it better, let us start by considering the case $J_1=0$.

\subsection{GSM degeneracy for $J_1=0$}

In this limit the system decomposes exactly into two independent spin rings, respectively made of the odd and the even lattice sites (see Fig.~\ref{fig:ANNNI_chain_decomposition}).
Both the two rings are made of an odd number of spins $L'=L/2=2N+1$ and are characterized by PBCs and AFM
nearest neighbor interactions.
Therefore, both rings are geometrically frustrated and admit $2L'=L$ independent ground states each~\cite{Maric2020_destroy, Odavic2023}. 
These states differ from each other in the position and in the orientation of a single ferromagnetic defect embedded in a Ne\'{e}l AFM state.
For each ring, they can be written as
\begin{equation}\label{kink_state_basis}
    \ket{k,\pm}_{o,e} = T_{o,e}^{k-1} \bigotimes_{j=1}^{L'-1} \sigma_{2j}^z \ket{\pm}_{o,e}^{\otimes L'}, 
\end{equation}
where $k$ runs from $1$ to $L'$ and marks the position of the ferromagnetic bond, $\ket{\pm}_{o,e}$ are the eigenstates of $\sigma^{x}_i$ on each odd/even ring, and $T_{o,e}$ is the translational operator acting on each of the two rings. As long as we keep $J_1=0$, each state constructed as the direct product of states in Eq.~\eqref{kink_state_basis} is a ground state of the whole system, resulting into a GSM with $L^2$ independent elements:
\begin{equation}
\label{eq:GSM0}
\mathcal{G}_0=\{\ket{k, \sigma}_{o}\ket{p,\sigma'}_{e}, k,p=1,\ldots L', \sigma,\sigma'=\pm\}.
\end{equation}

\subsection{GSM degeneracy for $J_1>0$}

For finite $J_1$, $\mathcal{G}_0$ splits into two subsets, each containing $L^2/2$ elements.
Which of the two has lower energy depends on the sign of $J_1$, and thus for definiteness let us focus on the AFM next-neighbor case ($J_1>0$ ). 
As shown in Fig.~\ref{fig:config}, it is clear that once a ferromagnetic defect is placed in one of the rings, the spin in the other ring lying between the two aligned spins minimizes the $J_1$ interaction by pointing in the opposite direction, thus halving the lowest energy configurations.

Therefore,for $J_1>0$, $h=0$ and $\kappa>1/2$ the GSM becomes
\begin{eqnarray}\label{GSM}
    \mathcal{G} & = & \!\! \left\{ \ket{\psi(k,p)}\equiv\ket{k,(-1)^{k}}_o \ket{p, (-1)^{p+1}}_e,\right. \nonumber \\
    & & \left. k= 1,\ldots, L, \;\;\;  p=k,\ldots,\! k\!+\!L'\!-\!1 \right\},
\end{eqnarray}
where we exploit the PBCs of each ring, i.e. $k,p\equiv k,p \mod L'$, and with a slight abuse of notation we identify the two eigenstates of $\sigma^x$ with their eigenvalues, i.e. $\pm=\pm 1$.
To make the relationship between the elements of $\mathcal{G}_0$ and the ones of $\mathcal{G}$ as clear as possible, in Fig.~\ref{fig:main_ANNNI_table} we provide a simple pictorial representation for the case $L=10$.


\usetikzlibrary{arrows.meta}

\begin{figure}[t]
    \begin{tikzpicture}[
        node distance = 0mm and 9mm,
        C/.style = {circle, draw, minimum size =6mm, inner sep=0pt},
        arr/.style = {-Straight Barb, semithick} 
    ]
    \node (abis) [] at (-0.1,1.5) {$\ldots$};
    \node (a) [C] at (0,0) {};
    \node (b) [C] at (1.5,0)  {};
    \node (c) [C] at (3,0)  {$+$};
    \node (d) [C] at (4.5,0)  {$+$};
    \node (e) [C] at (6,0)  {};
    \node (f) [C] at (.75,1.5)  {};
    \node (g) [C] at (2.25,1.5)  {};
    \node (h) [C] at (3.75,1.5) {$-$};
    \node (i) [C] at (5.25,1.5)  {};
    \node (l) [C] at (6.75,1.5)  {};
    \node (ebis) [] at (6.75,0.0) {$\ldots$};


    \foreach \n/\pos in {k/c}
        \node[below=1.5mm] at (\pos.west) {\n};

    \foreach \n/\pos in {k+1/d}
        \node[below=1.5mm] at (\pos.east) {\n};

    \foreach \n/\pos in {k/h}
        \node[above=1.5mm] at (\pos.west) {\n};

    \foreach \n/\pos in {odd chain/h}
        \node[below=15.5mm] at (\pos.south) {\n};

    \foreach \n/\pos in {even chain/h}
        \node[above=1.mm] at (\pos.north) {\n};
    

    \draw[]  (a) -- (b);
    \draw[]  (b) -- (c);
    \draw[red,line width=2pt]  (c) -- node[above=1mm,black]{$J_2$} (d);
    \draw[]  (d) -- (e);
    \draw[]  (f) -- (g);
    \draw[]  (g) -- (h);
    \draw[]  (h) -- (i);
    \draw[]  (i) -- (l);
    \draw[dashed]  (a) -- (f);
    \draw[dashed]  (b) -- (f);
    \draw[dashed]  (b) -- (g);
    \draw[dashed]  (g) -- (c);
    \draw[dashed,red,line width=2pt]  (h) -- node[left=.25mm,black]{$J_1$} (c);
    \draw[dashed,red,line width=2pt]  (h) -- (d);
    \draw[dashed]  (i) -- (d);
    \draw[dashed]  (i) -- (e);
    \draw[dashed]  (l) -- (e);
    \end{tikzpicture}
    \caption{Specifying the configuration of the odd chain, namely fixing the ferromagnetic defect position ($k$ in the figure), constrains the
    site on the even chain coupled to $k$ and $k+1$ by $J_1$ to have the opposite orientation to lower the state's energy.
    The case of a state of type $\ket{k,+}_o$ is illustrated. The overall effect of the $J_1$ interaction is then to split the degeneracy and  reduce the number of low energy states on the even chain from $L^2$ to $L^2/2$.}
    \label{fig:config}
\end{figure}
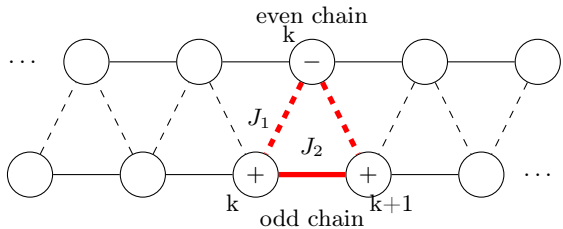

\subsection{Ground state near the classical point}

A small transverse magnetic field $h$ further breaks the super-extensive degeneracy left in $\mathcal{G}$, producing a band of closely lying states with a unique ground-state $\ket{g}$, similarly to what occurs for 1D chains with only next-neighbor interactions~\cite{Sen2008, Maric2022_fate, Catalano2022}. To extract the ground-state of the system we employed lowest order degenerate perturbation theory.
Despite the intricate structure of the perturbation matrix resulting from the interaction between the two TF rings, using graph theory (details for this calculation are presented in Appendix~\ref{sec:ground_statte_finding}), we are able to recover the analytic expression for the ground state in the antiphase with $L=4n+2$ sites close to the classical line for  $J_1>0$. It reads
\begin{equation}\label{GS}
    \ket{g} = \!A\sum_{k=1}^{L}\sum_{p=k}^{k+L'-1} \!\! \sin\left[\frac{(p-k+1) L \pi}{L+2}\right] \ket{\psi(k,p)},
\end{equation}
where \mbox{$A=2/\sqrt{L\left(L+2\right)}$} is the normalization constant. 

From Eq.~\eqref{GS} we can recover the physical quantities of interest, but we will focus in particular on the bipartite Von Neumann Entanglement Entropy~\cite{VN1, VN2}.

\subsection{EE near the classical point}

\begin{figure}[t]
	\includegraphics[width=1.0\linewidth]{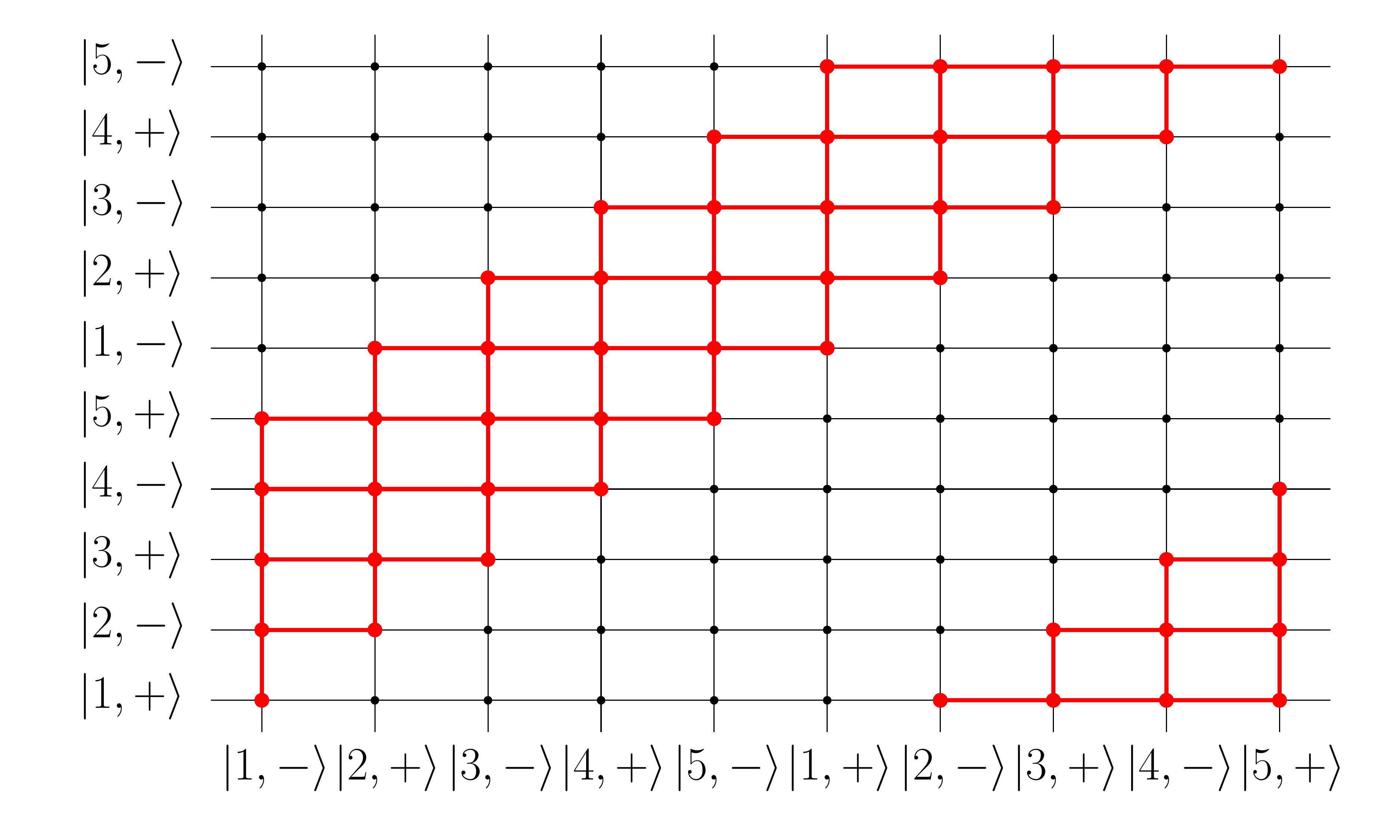}
	\caption{Pictorial representation of $\mathcal{G}_0$ and $\mathcal{G}$ for a system of length $L=10$. 
		The states of the even and odd rings are represented respectively on the horizontal and vertical axis. 
		Every vertex in the periodic grid is an element of the set $\mathcal{G}_0$. The effect of the AFM interaction ($J_1>0$) is to select the subset compatible with the constraint in Fig.~\ref{fig:config}, which are represented by the red vertices.}
	\label{fig:main_ANNNI_table}
\end{figure}

The EE  for the ground state $\ket{g}$ in Eq.~\eqref{GS} with respect to  a bipartition of the chain into a subsystem $A$ made by $M$ contiguous spins and its complement $\bar{A}$ is given by
\begin{equation}
\label{eq:EE}
S_M(\rho_A)=\mathrm{Tr}\left[\rho_A\log\rho_A\right].
\end{equation}
Here $\rho_A$ is the reduced density matrix obtained by tracing out from $\ket{g}$ all degrees of freedom outside $A$, i.e. $\rho_A = \Tr_{\bar{A} }(\ket{g}\bra{g})$. To evaluate the EE we have to determine the spectrum of $\rho_A$. Let us a sketch of the calculation here, while the details can be found in Appendix~\ref{Asym_EE}.

\begin{figure}
    \includegraphics[width=0.9\linewidth]{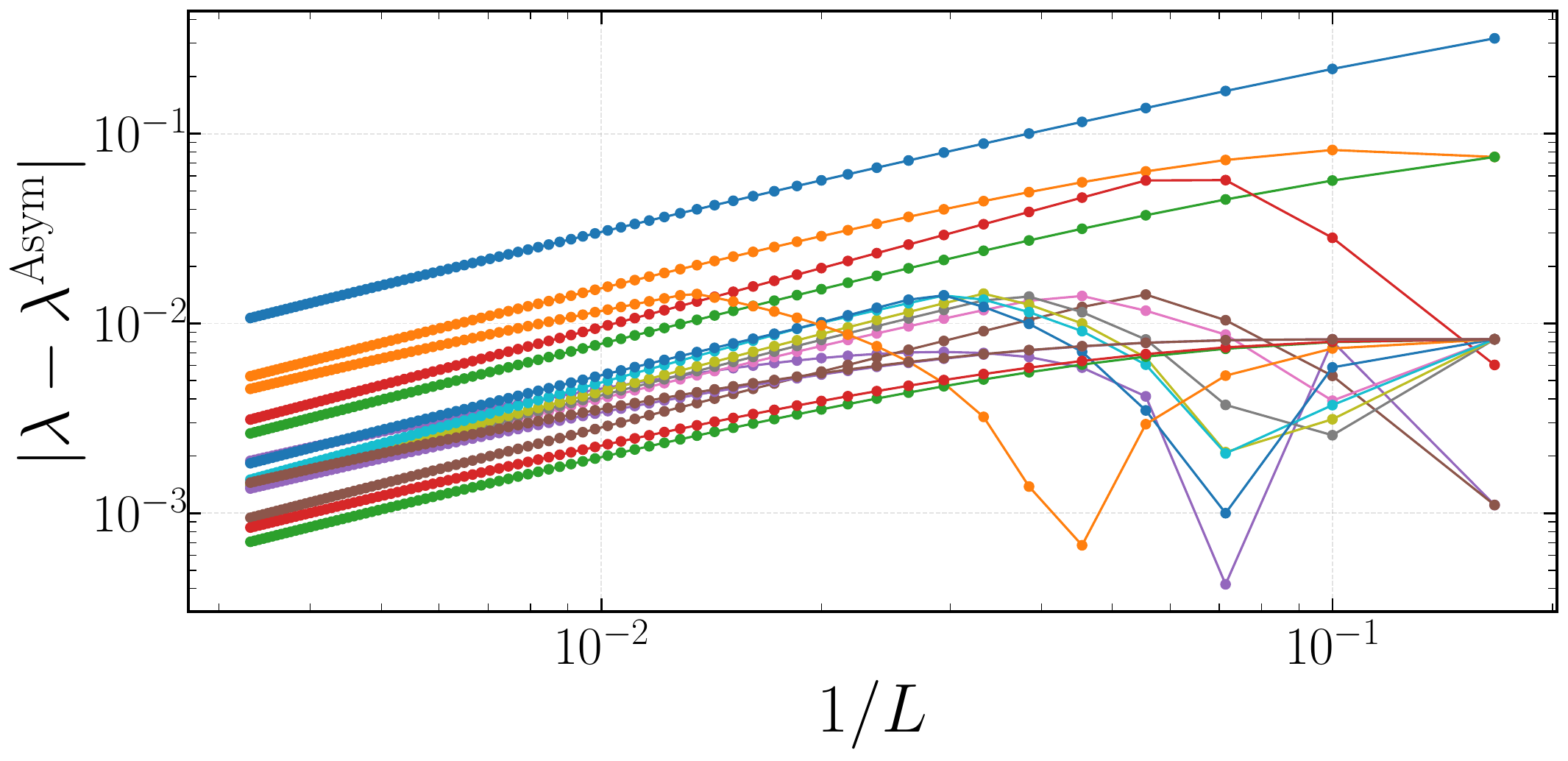}
    \caption{Absolute values of the difference of the numerical obtained $16$ non-zero eigenvalues of $\rho_A$ 
    and the asymptotic values in Eq.~\eqref{eig_limit} as function of the inverse of the chain length $L$. }
    \label{fig:main_density_matrix_approximation}
\end{figure}

We start noticing that, if both $A$ and $\bar{A}$ are made of more than four sites, only 16 eigenvalues of $\rho_A$ are different from zero (and appear in 4 degenerate multiplets with multiplicity four), while all the others vanish identically. Taking into account the interaction graph showed in Fig.~\ref{fig:main_ANNNI_table}, it is possible to re-order the basis elements according to the position of the ferromagnetic defects being inside or outside $A$, such that a block-structure emerges in $\rho_A$.
Although the resulting reduced density matrix is not block-diagonal, it is possible to prove that the off-diagonal block provide only subleading corrections to the matrix eigenvalues and thus in the thermodynamic limit the eigenvalues of $\rho_A$ coincides with those of the diagonal blocks, which are
\begin{eqnarray}\label{eig_limit}
    \lambda_{1} (x) & = & \dfrac{(1-x)^2}{4}- \dfrac{\sin^2\pi x}{4\pi^2}, \nonumber \\
    \lambda_{2} (x) & = & \dfrac{x^2}{4}-\dfrac{\sin^2\pi x}{4\pi^2}, \\
    \lambda_{3,4}(x) & = & \dfrac{x(1-x)}{4} + \dfrac{\sin^2\pi x}{4\pi^2} \pm \dfrac{\sin\pi x}{4 \pi}, \nonumber
\end{eqnarray}
where $x=M/L$ is the relative dimension of the partitions with respect to the chain length.
In Fig.~\ref{fig:main_density_matrix_approximation} we can observe for $M=L/2$ how the eigenvalues of $\rho_A$ obtained from numerical diagonalization tend to coincide with the analytically determined values in the thermodynamic limit.
Similar results can be obtained for different values of $x$, hence proving the validity of Eq.~\eqref{eig_limit}.

From Eq.~\eqref{eig_limit} it is possible to recover the expression of the entanglement entropy for diverging $L$ when $J_1\neq0$. 
Remarkably, it can be put in the form
\begin{eqnarray}\label{EE_anal_result}
    S_M(\rho_A) & = & 2 - y\log_2{y} - (1-y)\log_2{(1-y)} \nonumber \\
    & & - z\log_2{z} - (1-z)\log_2{(1-z)},
\end{eqnarray}
where 
\begin{equation}
 y=x-\frac{\sin\pi x}{\pi}, \qquad {\rm and} \qquad z=x+\frac{\sin\pi x}{\pi}.
 \label{yzdef}
\end{equation}

The first term on the RHS of Eq.~\eqref{EE_anal_result} stems from the four-fold degeneracy of the eigenvalues and is due to the double $\mathcal{Z}_2$ symmetry of flipping each spin on each even/odd sub-lattice. The other terms represent the EE of two delocalized particles, each of them having probability $y$ ($z$) of being in $A$.
The factorization of the entropy contribution for each excitation indicates their independence, however their coefficients do not coincide with the geometrical probabilities $y=z=x = M/L$ (which emerge for $J_1=0$): the corrections in eq.~\eqref{yzdef} indicate a correlation between these excitation.
Indeed, they are consistent with the result in Ref.~\cite{Berkovits2013}, where the EE was calculated in the case of few excitations over the vacuum of a quadratic theory. In that case, each particle contributes with a probability that displays a correction due to the relative distance in momentum space. Eq.~\eqref{yzdef} fits the results in ref.~\cite{Berkovits2013} if the two excitations in the ANNNI chain differ by $\Delta k = \frac{2\pi}{L}$, indicating that they are trying to minimize their kinetic energy under the Pauli-like constraint of occupying different momentum states.

Thus, quite remarkably, we found that the ground state EE of the TF ANNNI is compatible with the existence of two excitations. 
While the effects of geometrical frustration are commonly interpreted through a single particle description, the existence of multiple proper quasi-particles due to more complex frustration is quite unexpected. Based on previous results~\cite{Maric2022_fate,Catalano2022,Maric2020_destroy,Maric2020_neworder}, we expect that the emergence of this quasi-particle picture will have consequences also on other observables of the system. 
This is the case, for example, for the energy gap, which is expected to close as $L^{-2}$ in a TF system, or for order parameters in the antiphase, which are expected to be destroyed or become incommensurate. 
Since in this work we focus on the EE properties, we leave a detailed analysis of these features to a future work.

\section{Numerical analysis}
\label{num_analys}

\begin{figure}
    \includegraphics[width=.99\columnwidth]{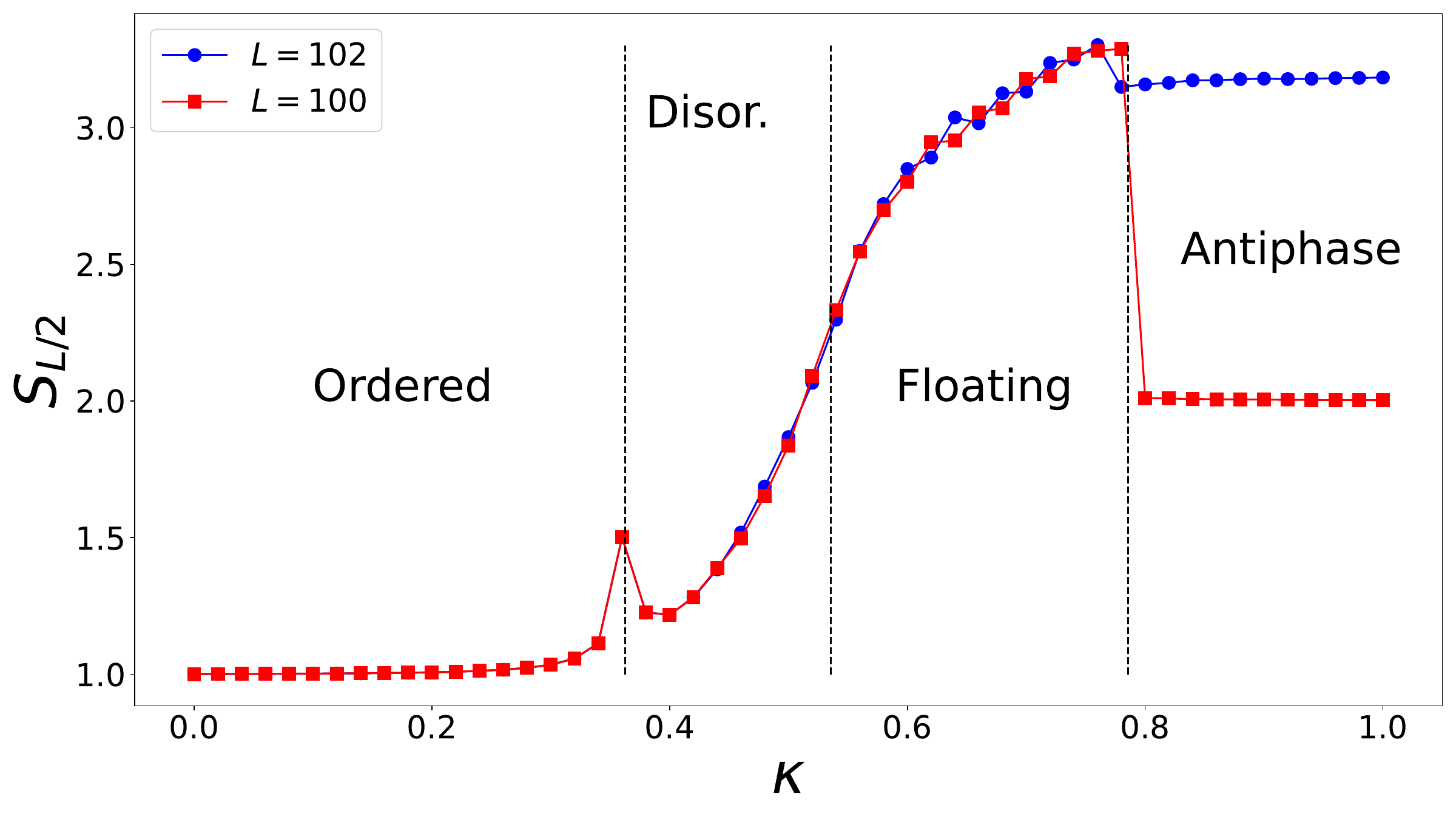}
    \caption{Comparison of the EE at half chain as a function of $\kappa$ for $h/J_1=0.3$, between  the case with TF ($L=4N+2$ blue dots), and without TF ($L=4N$ red squares). 
    The numerical data are obtained setting $N=25$.}
    \label{fig:PhaseDiagramEnt}
\end{figure}

The ANNNI model Eq.~\eqref{eq:Hamiltonian} is notoriously not analytically solvable: to obtain results beyond the perturbative regime employed so far we use a density matrix renormalization group algorithm (DMRG) based on tensor networks~\cite{White92, Orus14, Orus19, Biamonte20, Catarina23}, in which the ground state is represented through a matrix product state (MPS). 
In order to avoid ambiguities in the canonization of the MPSs, we implemented the periodic boundary conditions directly in the matrix product operator (MPO) encoding the system's Hamiltonian~\cite{Weyrauch2013} rather then using periodic MPSs~\cite{Verstraete2004,Pippan2010}.
Some details about our numerical approach can be found in Appendix~\ref{ap:Numerics}. 
The MPS approximation is known to be efficient for states possessing a finite amount of entanglement, as in the case of the ground-states of one-dimensional systems satisfying the area law~\cite{Vidal2003, Latorre2004, Eisert2010} and has been already applied successfully for the ANNNI chain~\cite{Beccaria06, Beccaria07}. 
From the results of the previous section, and in analogy with the ones obtained in other systems with TF~\cite{Giampaolo2019}, we expect that
the amount of entanglement in the ground state 
must stay finite inside the antiphase even in presence of TF.
Hence, the MPS representation of the ground state must be a faithful one.
Such a representation of the ground state is particularly useful to evaluate the EE since the extraction of the partition related Schmidt coefficients~\cite{VN2} is straightforward.

Let us start with an overview of the general behavior of the EE across the phase diagram: we set the length of $A$ to a half chain ($M=L/2$) and we evaluate the EE as a function of $\kappa$ for $h/J_1=0.3$. The results obtained both for $L=4N$ and $L=4N+2$ with $N=25$, are plot in Fig.~\ref{fig:PhaseDiagramEnt}. In both the ordered and disordered phases,
the EE values are practically indistinguishable from each other. Even in the floating phase, they tend to coincide in the limit of large $N$, although the convergence is slower and for $N=25$ the differences can still be observed. This slow convergence is not a surprise, since the floating phase is known to be gapless and well approximated by a conformal field theory and thus the EE at half chain keeps growing logarithmically with the chain length and finite size effect are more prominent~\cite{Suzuki2013}.

\begin{figure}[t]
	\includegraphics[width=.99\columnwidth]{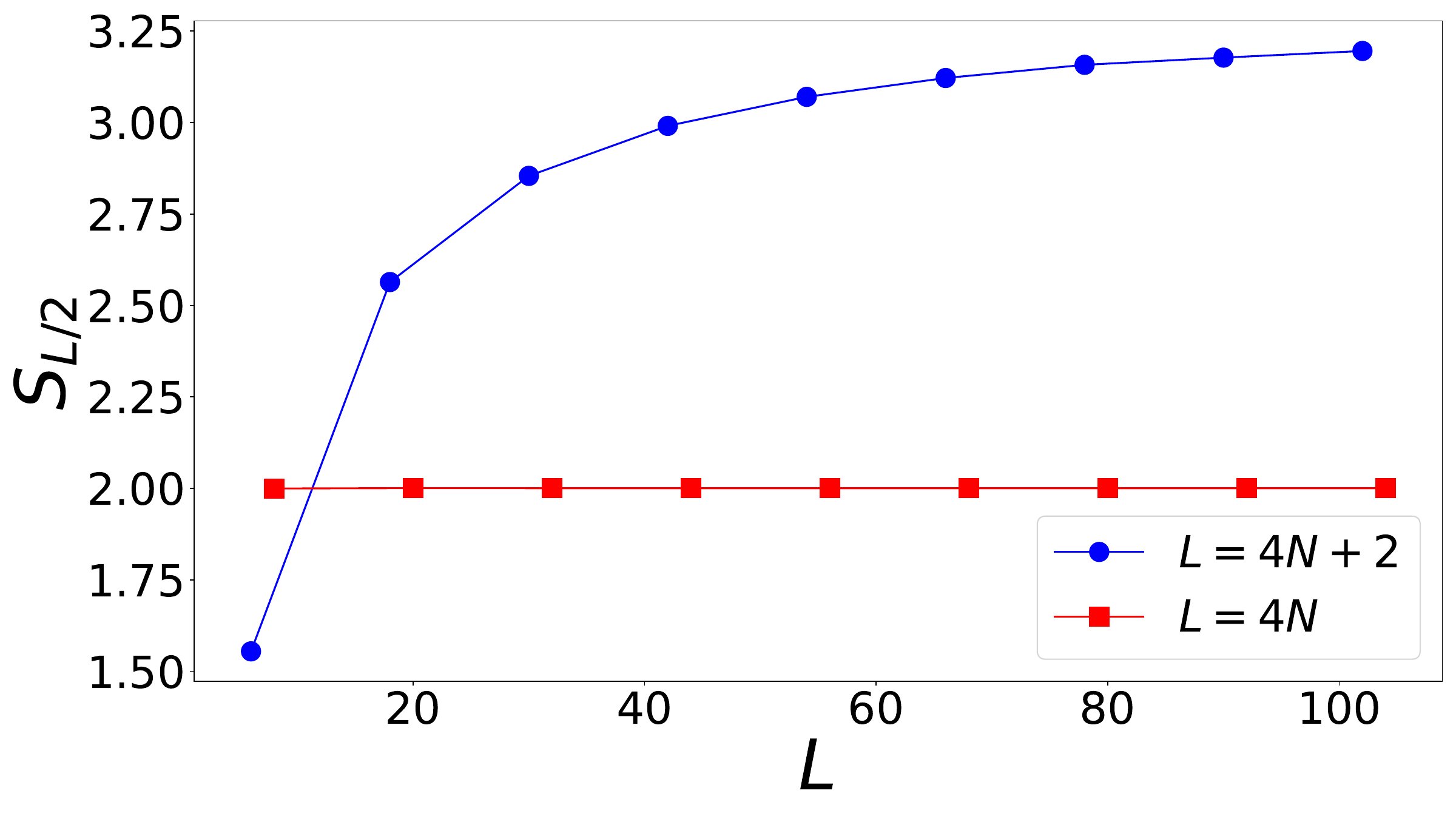}
	\caption{Bipartite Entanglement Entropy at half chain as a function of the length of the chain. The data are obtained using our DMRG algorithm for $J_2=1$, $h=0.2$, $J_1=1$, for the case with TF ($L=4N+2$ blue dots), and without TF ($L=4N$ red squares).}
	\label{fig:HalfChain}
\end{figure}

\begin{figure}[t!]
	\centering
	\includegraphics[width=.99\columnwidth]{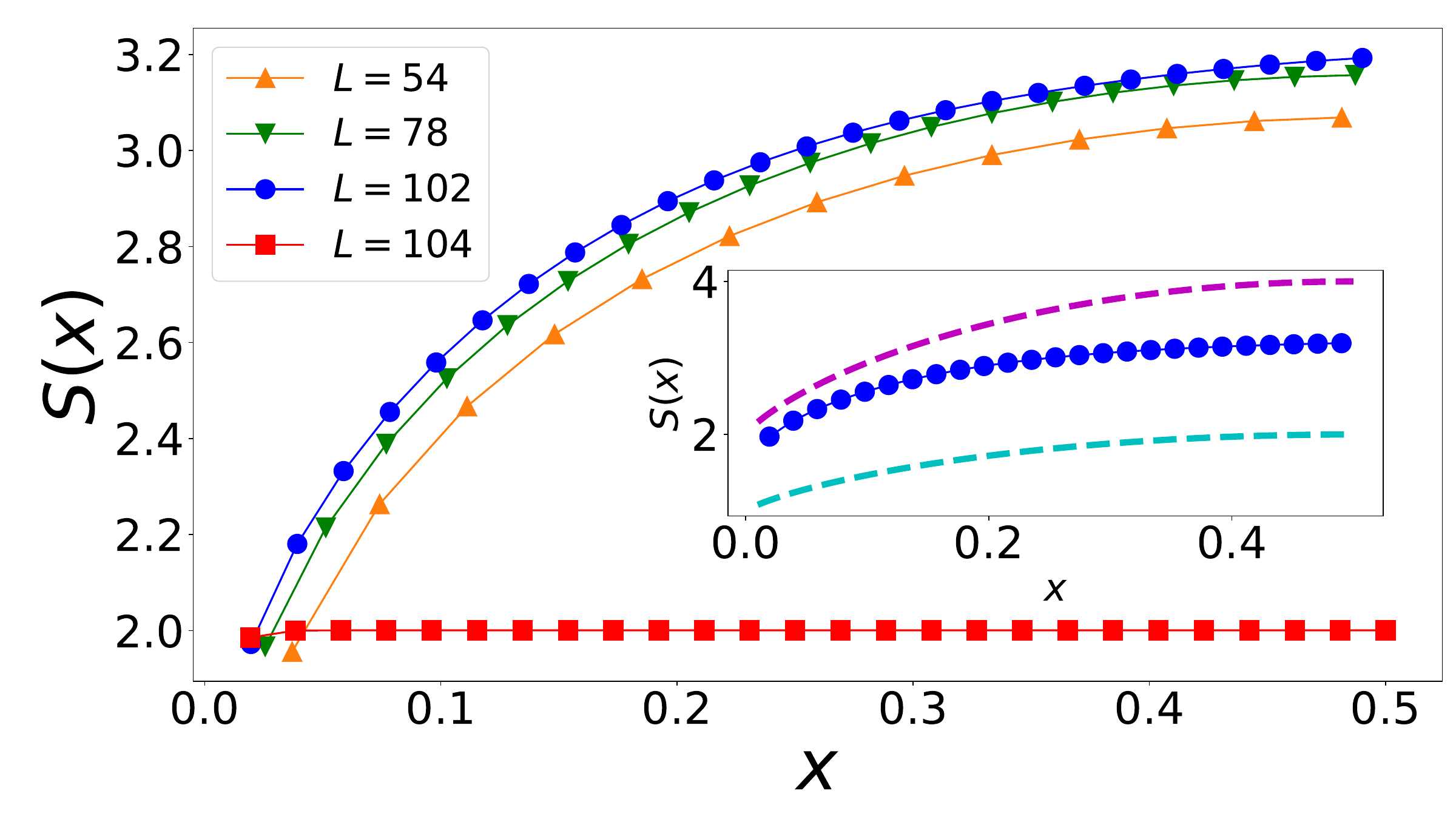}
	\caption{Bipartite entanglement entropy as a function of the ratio $x=M/L$ between the subsystem length and the length of the chain. The data is obtained using our DMRG algorithm for $J_2=1$, $h=0.2$, $J_1=1$, for the TF (blue dots, green and orange triangles) and locally frustrated (red squares) ANNNI model, with $L=102$, $78$, $54$ and $L=104$ respectively. Inset: EE of the TF chain for $L=102$ compared to the EE for a state with one (cyan dashed line) and two (purple dashed line) excitations (eq.~\eqref{EE_anal_result} with $y=x$,$z=0$ and $y=z=x$ respectively).}
	\label{fig:EE}
\end{figure}

On the contrary, the behavior of the EE becomes different as soon as we enter the antiphase, where the topological frustration drastically affects the behavior of the entanglement. Such a difference cannot be explained as finite size effects, as evidenced  in Fig.~\ref{fig:HalfChain}, where the dependence of the EE at half chain is analyzed by varying the chain length for a fixed set of the Hamiltonian parameters inside the antiphase. While for systems whose length is an integer multiple of four the EE is virtually independent from the size, TF
induces a dependence on $L$ in the EE that, however, remains finite also in the thermodynamic limit. Fig.~\ref{fig:EE} highlights the different subsystem dependence of the EE in the two cases, with the not TF case quickly saturating to the (constant) area law. The inset further show that in the TF case the growth of the EE with the subsystem size lies in between that expected for a single and two delocalized, independent excitations. 

\begin{figure}[t!]
    \includegraphics[width=.99\columnwidth]{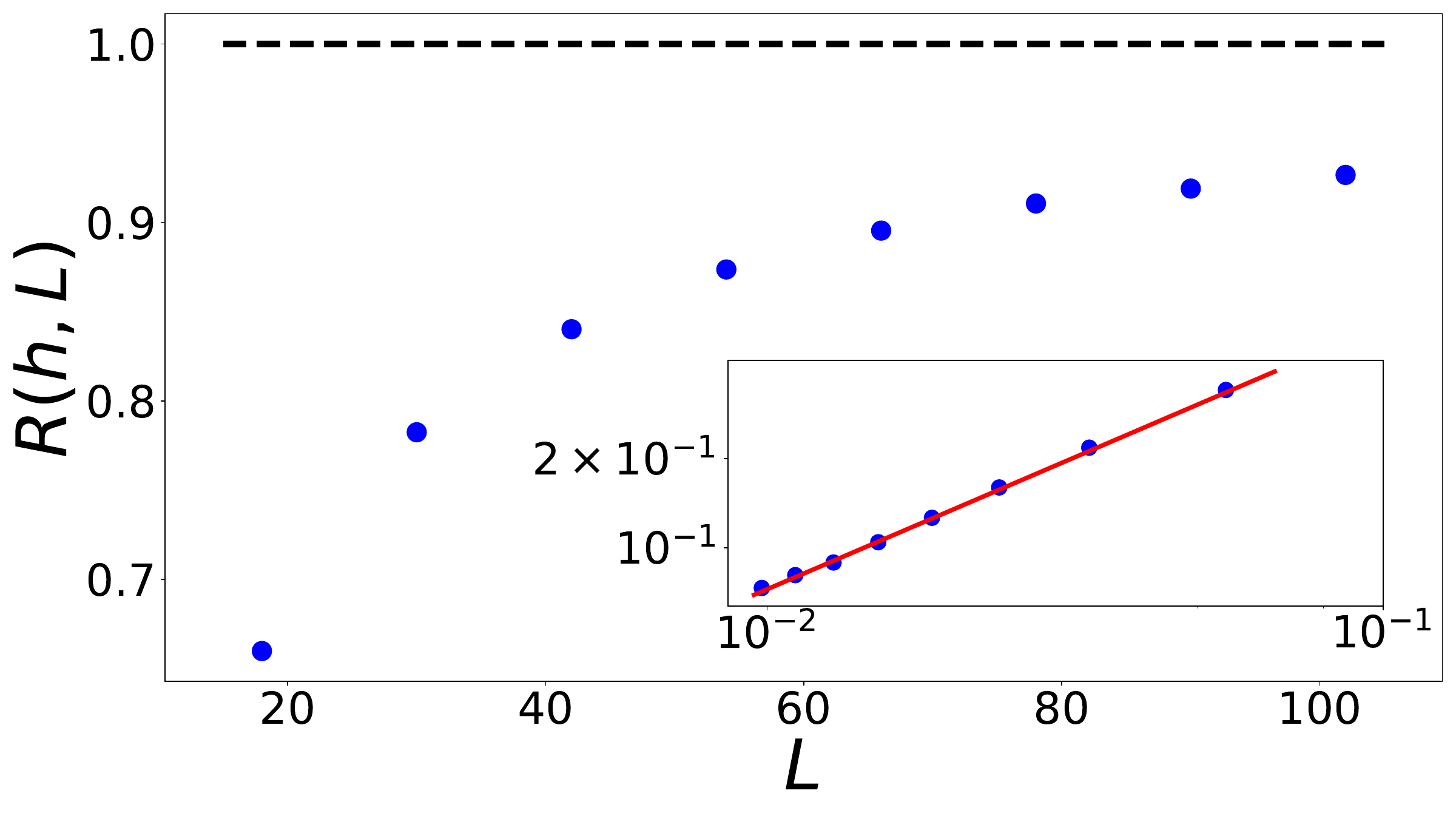}
    \caption{$R(h,L)$, defined in Eq.~\eqref{Rdef}, as a function of the inverse system's size $L^{-1}$ for $h=0.2$, $\kappa=1$. The inset shows a plot in log-log scale of $1-R$ vs $L$, together with a linear fit, hinting to a power-law convergence in the thermodynamic limit, describe by the numerical law $\log(1-R)=-0.89\log L + 1.47$.}
    \label{fig:EE_sum}
\end{figure}

From previous results on other TF models, a noteworthy observation emerged: in the thermodynamic limit, the different contributions to quantum resources decouple in the sum of two terms in which one coincide with the resource of the unfrustrated counterpart and the other is purely due to TF~\cite{Giampaolo2019, Odavic2023}.
This naturally leads to the hypothesis that a similar behavior occurs also in the EE of the ANNNI model.
However, providing evidences that support this hypothesis proves to be more delicate than in the previous cases, since for the ANNNI isolating the local contribution is challenging. In models with only nearest neighbor coupling, it is enough to change the interaction sign, thus removing TF. In the ANNNI model, reversing the signs of the interactions eliminates both TF and also the local frustration. 
Therefore, to remove TF while preserving the other local properties, we evaluate the EE for the same set of the Hamiltonian parameters $\mathcal{J} \equiv (J_1,J_2,h)$ and with the same length $L$ but applying open boundary conditions and considering the subset $A$ in the middle of an open chain, to reduce boundary effects. 
We compare the EE obtained in this way, which we denote as $S_{M}^{o}(\mathcal{J},L)$, with the EE obtained assuming PBCs, namely $S_{M}^{p}(\mathcal{J},L)$.
If the hypothesis stands, the difference of these two quantities must be equal to the topological contribution that is provided by 
Eq.~\eqref{EE_anal_result}.
In other words, if the hypothesis is verified the quantity 
$R(\mathcal{J},x,L)$, defined as
\begin{equation}
	\label{Rdef}
   R(\mathcal{J},x,L)=\frac{S_{M}^{p}(\mathcal{J},L)-S_{M}^{o}(\mathcal{J},L)}{S_M(g)},
\end{equation}
when $L\rightarrow\infty$, shall tend to $R(\mathcal{J},x,L)\rightarrow 1$. 
The data depicted in Fig.~\ref{fig:EE_sum} clearly support this hypothesis. 
The analysis carried out in the inset indicates a power-law convergence in the thermodynamic limit with $\log(1-R)=-0.89\log L + 1.47$.

It is worth mentioning here that, we numerically observe that the local contribution to the EE in addition to the global $2$ factor is minimal (approximately $10^{-3}$) within the bulk of the antiphase and increases only close to the phase transition. Hence, even at finite $h$, the total entanglement entropy of the frustrated system can be well approximated by the perturbative formula obtained for $h \ll J_1$.


\section{Conclusion and  outlook}\label{conclusion}

In summary, we considered the effects of the interplay between local and non-local sources of frustration in 1D spin chains through the analysis of the entanglement entropy of the ANNNI model. The first type of frustration is due to the interplay between the nearest and next-to nearest neighbor interactions, while the non-local source (TF) is injected by a suitable choice of boundary conditions.
In this work our focus centered on the antiphase region, where we revealed the presence of TF even in systems with an even number of spins, when the number of sites is not divisible by four. In this way, we showed that TF can emerge beyond the usual framework of geometrical frustration induced by FBCs (odd number of sites), presenting both usual and novel features.

The emergence of TF with an even number of sites can be traced back to the fact that, in the limit in which the interaction with first neighbors becomes negligible, the system decomposes into two independent topologically frustrated quantum Ising models. 

Starting from this consideration, exploiting an analytical, perturbative approach valid close to the classical limit, we managed to obtain an analytical expression of the unique ground state of the model. From it, we were able to obtain the value of the EE for a generic bipartition composed of connected subsets of contiguous spins in the thermodynamic limit. 
We proved that the EE is decomposed into a contribution present also without TF and one due to TF that indicates that the ground state hosts two excitations. The latter terms are independent, except for the fact that both particles tend to minimize their momenta, but cannot occupy the same momentum state and in this way they develop a correlation, as often happens in 1D system, where excitations typically acquire a fermionic nature.

To study the EE in the whole phase diagram of the model we employed a tensor network based DMRG code and observed that for $L=4N+2$ the effect of TF extends, and is limited, to the entire antiphase. There, we showed that the EE can be decomposed again into a non TF contribution (obtained by applying open boundary conditions to the same Hamiltonian) and that of two excitations, with the same values obtained in the analytical perturbative regime.
The importance of this results stems from its consistency with what happens to every quantum resource analyzed so far in models with only nearest-neighbor interaction. It therefore supports the idea that this decomposition of quantum resources is a general characteristic of topologically frustrated systems.

We would like to stress once more that the unveiled quasi-particle description of the TF ANNNI chain is unexpected and required a meticulous effort to be exposed. We expect that similar results may be unveiled also for other frustrated systems. We plan to continue this investigation and in the next work, we will consider the same ANNNI model, but with an odd number of sites. There the interplay between local and topological frustrations is even more intricate and affects the model beyond just the antiphase.

In general, we showed that the (extensive) frustration emerging from the competition between local interactions and the intensive one of topological origin contribute differently to the phenomenology of the model. In the future we will consider other properties, such as order parameters, complexity, quantum coherence, etc., but even more interesting will be to explore other models with different types of frustration to map and understand the phenomenology of the interplay between various source of frustration, eventually moving to higher dimensions as well.

\begin{acknowledgments}
AGC acknowledges support from the MOQS ITN programme, a European Union’s Horizon 2020 research and innovation program under the Marie Sk\l{}odowska-Curie grant agreement number 955479. 
SBK is supported by the Croatian Science Foundation (HrZZ) Project DOK–2019–4–3321
\end{acknowledgments}

\appendix

\section{Perturbation theory close to the classical line}
\label{sec:app_analytical}

In this appendix we show the details of how to obtain the ground state of the topologically frustrated ANNNI model near the classical line, i.e. in the limit $h\rightarrow 0^+ $, resorting to the lowest order perturbation theory and how to use this result to extract its bipartite entanglement entropy.

\subsection{Determination of the ground state}
\label{sec:ground_statte_finding}

Let us divide the full Hamiltonian in eq.~\eqref{eq:Hamiltonian} as
\begin{equation}
    H=H_0+h\,\mathcal{H}^{\textrm{pert}},
\end{equation}
with 
\begin{equation}
    \label{eq:H0}
    H_0=J_1\sum_{i=1}^L \sigma_{i}^x\sigma_{i+1}^x + J_2\sum_{i=1}^L \sigma_{i}^x\sigma_{i+2}^x,
\end{equation}
and 
\begin{equation}\label{H_Int}
    \mathcal{H}^{\textrm{pert}} = \sum_{j=1}^L \sigma_j^x = \mathcal{H}^{\textrm{pert}}_{\textrm{o}} + \mathcal{H}^{\textrm{pert}}_{\textrm{e}}, \quad \mathcal{H}^{\textrm{pert}}_{\textrm{o}/\textrm{e}} =  \sum_{\substack{j \in \{o/e\}}}\sigma_j^x,
\end{equation}
where, keeping the notation introduced in the main text, the \textit{e} (\textit{o}) subscript refer to the even (odd) sites subchain.

\begin{figure}[t!]
	\includegraphics[width=0.9\linewidth]{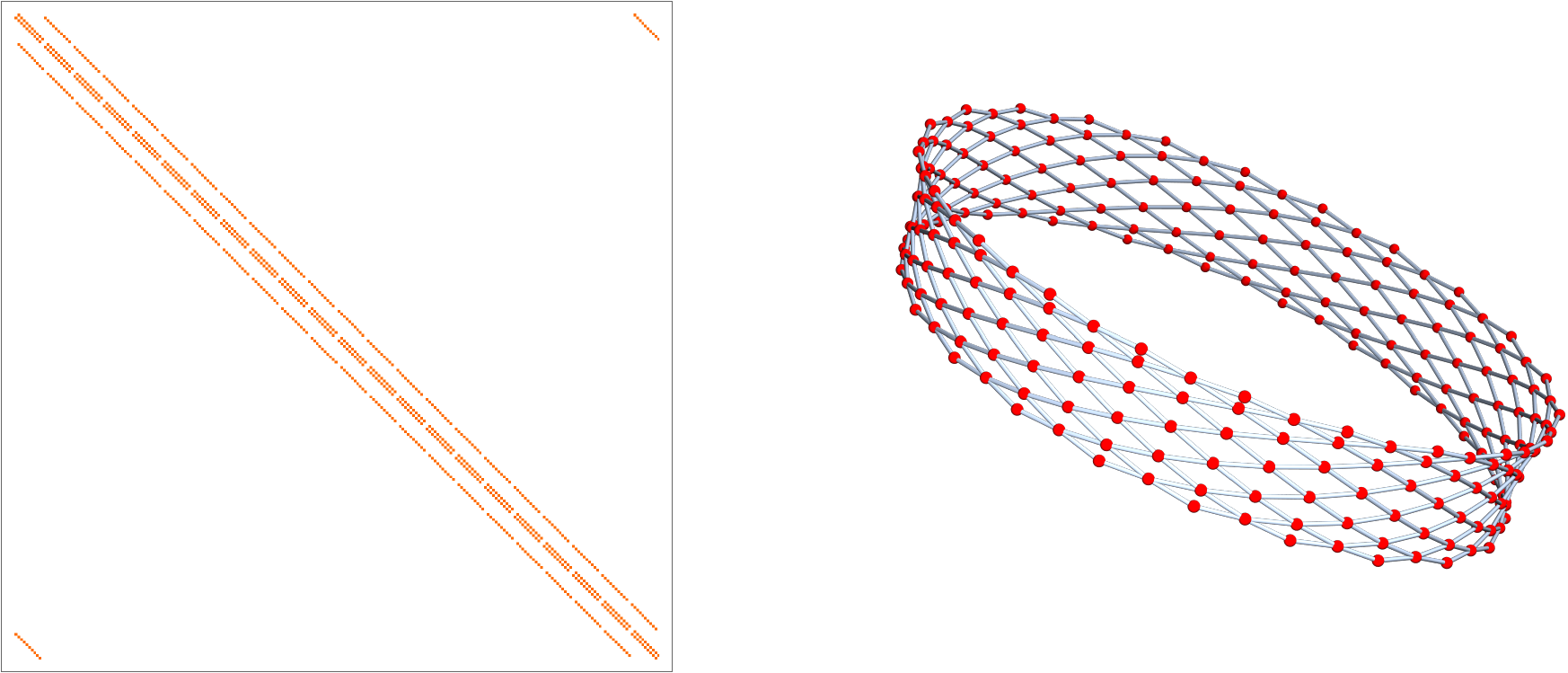}
	\caption{\label{fig:matrix_graph_L_14}Left: Matrix representation of the perturbation Eq.~\eqref{H_Int} over the set $\mathcal{G}$ given in Eq.~\eqref{GSM} for the ANNNI chain of length $L=14$. Right: Graph representation of the matrix on the left.}
\end{figure}

To find the analytical expression of the ground-state close to $h=0$ we thus need to diagonalize the matrix $\mathcal{H}^{\textrm{pert}}$ over the degenerate ground-state manifold $\mathcal{G}$ (Eq. \eqref{GSM} in the main text), which reads
\begin{align}
	&\bra{\psi_{(k,p)}} \mathcal{H}^{\textrm{pert}} \ket{\psi_{(k',p')}} = f_{\textrm{o}}(k,p,k',p') + f_{\textrm{e}}(k,p,k',p'), \label{H_pert_matrix_element} \\
	& f_{\textrm{o}}(k,p,k',p') \equiv \left( \delta_{k,k'-1}\delta_{(-1)^{k},(-1)^{k'-1}}+ \right. \nonumber \\
	& \left. + \delta_{k,k'+1}\delta_{(-1)^{k},(-1)^{k'+1}}\right) \times \delta_{p,p'} \delta_{(-1)^{p},(-1)^{p'+1}}, \nonumber \\
	& f_{\textrm{e}}(k,p,k',p') \equiv \delta_{k,k'} \delta_{(-1)^{k},(-1)^{k'}} \times \nonumber \\
	& \times \left( \delta_{p,p'+1}\delta_{(-1)^{p+1},(-1)^{p'+2}}+ \delta_{p,p'-1}\delta_{(-1)^{p+1},(-1)^{p'}}\right), \nonumber
\end{align}
where periodicity of the indices with respect to $L'$ is intended. The matrix plot and its associated graph are shown in Fig.~\ref{fig:matrix_graph_L_14}.

\begin{figure}
	\includegraphics[width=0.9\linewidth]{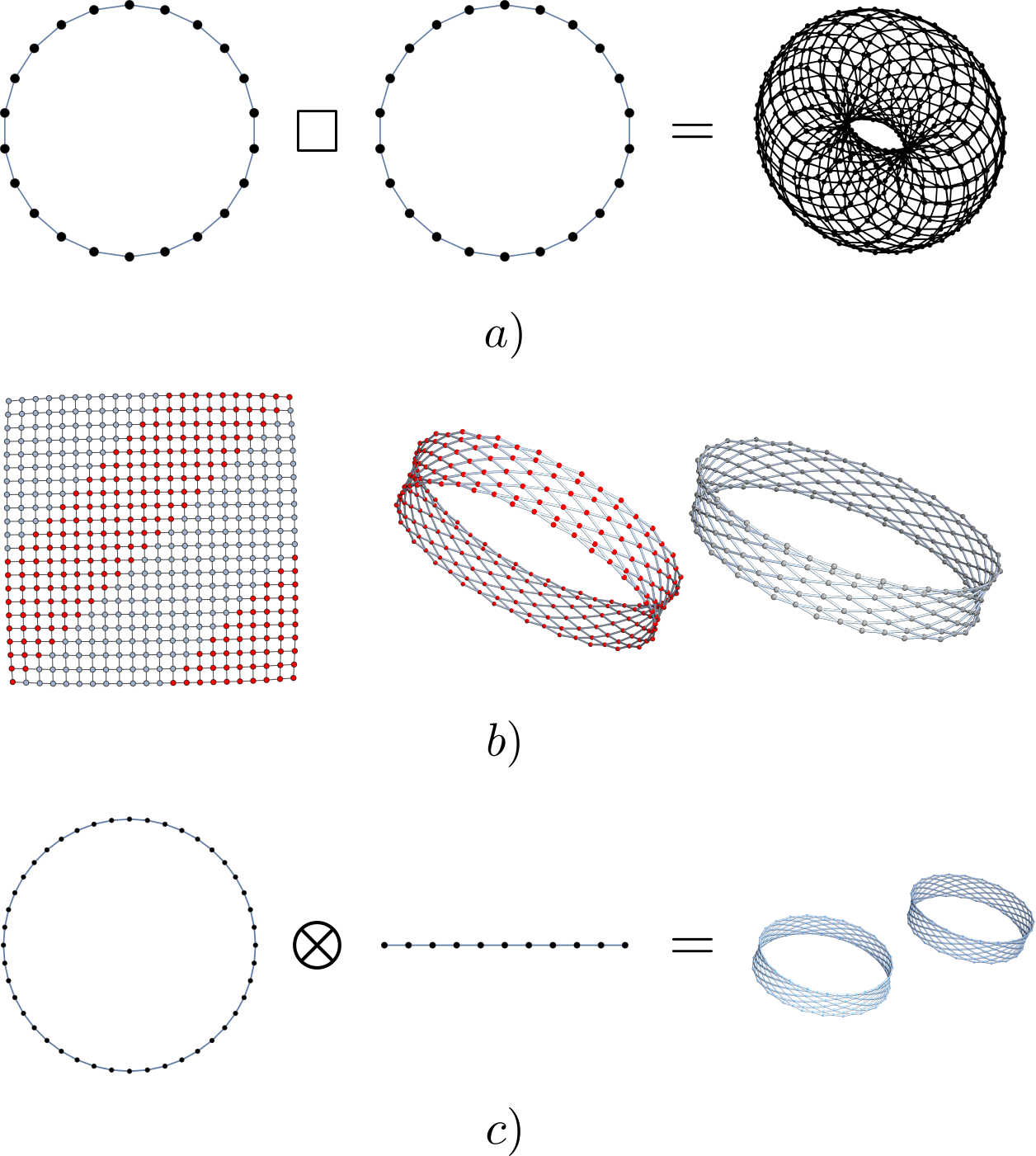}
	\caption{\label{fig:torus_graph_L_22}a) The Torus Grid graph $T_{L,L}$ obtained as the Cartesian product $C_L\square C_L$ for $L=22$. b) The unfolded Torus with the two subsets corresponding to $\mathcal{G}$ (red vertices) and $\mathcal{G}^c$ (gray vertices) highlighted, and the corresponding disconnected sub-graphs. c) The same sub-graphs can be also obtained as the tensor product of the Cycle graph of cardinality $2L$ and the Path graph of length $L'$.}
\end{figure}

Because of the complex structure of this matrix, its diagonalization is non trivial. Nevertheless, a better insight can be obtained by looking first at the simpler case in which $J_1=0$, corresponding to two non-interacting TF Ising rings. In this case, it is easy to check that $\mathcal{H}^{\textrm{pert}}$ can be written as the cartesian product (box product) of two circulant matrices when evaluated over the manifold $\mathcal{G}_0$ (see Eq.~\eqref{eq:GSM0} of the main text), namely 
\begin{equation}
    \mathcal{H}^{\textrm{pert}} = \mathcal{H}^{\textrm{pert}}_{\textrm{o}} \square \mathcal{H}^{\textrm{pert}}_{\textrm{e}}.
\end{equation}
While it is known that these matrices can be exactly diagonalized, it is worth noting that all the entries of $\mathcal{H}^{\textrm{pert}}_{o/e}$ are either zeros or one. Hence, they can be understood as adjacency matrices~\cite{lee93}, associated to two identical cycle graphs $C_\frac{L}{2}$~\cite{Clancy20}. Remembering that for any given two graphs $G_1$ and $G_2$ having adjacency matrix $A(G_i),\, i=1,2$ respectively, $A(G_{1}) \square A(G_{1}) = A(G_{2}\square G_{2})$~\cite{Barik18} (Fig,~\ref{fig:torus_graph_L_22}), we can conclude that the resulting graph associated with the full matrix $\mathcal{H}^{\textrm{pert}}$ is the Torus Grid Graph $T_{\frac{L}{2},\frac{L}{2}}=C_\frac{L}{2}\square C_\frac{L}{2}$~\cite{Clancy20}. 

As discussed in the main text, setting $J_1\neq 0$ acts as a selection rule: its effect being to split the set $\mathcal{G}_0$ in half according to the orientation of the spin of the second chain that falls between the two spins of the first ring where the magnetic defect is localized:
\begin{equation}
    \mathcal{G}_0 = \mathcal{G} \cup \mathcal{G}^c,
\end{equation}
where $\mathcal{G}^c$ is the complement set of $\mathcal{G}$. The graphs of the two sets $\mathcal{G}$ and $\mathcal{G}^c$ are respectively the red and gray part of the torus (see Fig.~\ref{fig:torus_graph_L_22}b for its unfolded representation and its decomposition). The red one is also the graph obtained through Eq.~\eqref{H_pert_matrix_element}, its ground state being then the required solution. We can now observe that this graph decomposition can be achieved through the Cartesian product of a Cycle graph of length $2L$ and a Path graph of length $L'$ (Fig.~\ref{fig:torus_graph_L_22}c). Specifically, the $L$ kink states of the odd chain couple separately with half of the kink states ($L'$) that are compatible with the condition $J_1>0$, and the other half with the condition $J_1<0$. The eigenvectors of the adjacency matrix of the desired graph (the red ring in Fig.~\ref{fig:torus_graph_L_22}) can then be obtained directly from the eigenvectors of the graphs involved in the operation (see Tab.~\ref{tab:graph_operation_properties}) through their tensor product, and finally projecting into the subspace of the desired ring, since the two graphs are disconnected (see Tab.~\ref{tab:graph_operation_properties}).

\begin{table}[t]
	\centering
	\begin{tabular}{ccc}
		\toprule
		\textbf{Graph} & \textbf{A-eigenvalues} & \textbf{A-eigenvectors} \\
		\midrule
		$G_1$ & $\lambda_i$ & $X_i$ \\
		\hline
		$G_2$ & $\mu_i$ & $Y_j$ \\
		\hline
		$G_1\square \, G_2$ & $\lambda_i + \mu_j$ & $X_i \otimes Y_j$ \\
		\hline
		$G_1\otimes \, G_2$ & $\lambda_i \mu_j$ & $X_i \otimes Y_j$ \\
		\hline
		\multirow{2}{*}{\parbox{2cm}{$G_1 \cup G_2$}} & $\lambda_i$ & $\left(\begin{array}{c}
			X_i\\
			0
		\end{array}
		\right)$ \\
		& $\mu_j$ &  $\left(\begin{array}{c}
			0 \\
			Y_j
		\end{array}
		\right)$ \\
		\hline
	\end{tabular}
	\caption{Properties of the graph operations considered. First column: graphs and graph operations. Second column: eigenvalues of the corresponding adjacency matrix $A$. Third column: Eigenvector of the corresponding adjacency matrix $A$.}
	\label{tab:graph_operation_properties}
\end{table}

\begin{table}[t]
	\centering
	\begin{tabular}{ccc}
		\toprule
		\textbf{Graph} & \textbf{A-eigenvalues} & \textbf{A-eigenvectors} \\
		\midrule
		\multirow{2}{*}{Cycle} & $\lambda_{l} = \sum_{k=0}^{m-1}c_k \omega^{kj}$ & \multirow{2}{*}{ $\ket{b_{l}} = \sum_{r=0}^{m-1}  e^{ i \frac{2 \pi}{m} l r }{}\ket{r}$} \\
		& $j=0,\ldots,m-1$ &  \\
		\midrule
		\multirow{2}{*}{Path} & $\mu_{k}=2\cos \left(\frac{k\pi}{m+1}\right)$ & \multirow{2}{*}{$\ket{a_{k}} = \sum_{j=1}^{m}\sin \left( \frac{k\pi}{m+1}j\right ) \ket{j}$} \\
		& $k=1,2,...,m$ &  \\
		\bottomrule
	\end{tabular}
	\caption{Eigenvalues and eigenvectors of the adjacency matrix of the Cycle and Path graphs having $m$ vertices. The coefficients $c_k$ is the matrix element on the $k$-th row of the Circulant matrix associated to the Cycle graph. }
	\label{tab: cycle_path_graphs_properties}
\end{table}

The general eigenvector will then be of the form $\ket{b_l}\otimes\ket{a_k}$, that are respectively eigenvectors of the Cycle and Path graphs, with eigenvalue $\lambda_l\,\mu_k$ (see Tab.~\ref{tab: cycle_path_graphs_properties}). The ground state will correspond to the $\{l,k\}$ values for which the product is minimum. Since the solely effect of the $J_1$ interaction is to select a sub-set of states, we expected for the odd chain the ground state of the topologically frustrated Ising ring ($l=0$). As a consequence, the product is minimized for $k=L'$, and the ground states will be of the form $\ket{b_0}\otimes\ket{a_{L'}}$. The ground state will then be of the form
\begin{equation}\label{temporary_gs}
    \ket{g} \! = \! \!A\sum_{k=1}^{L}\!\sum_{r=0}^{L'-1} \sin\left[\alpha(r+1)\right] \ket{k,(-1)^k}\ket{k+r,(-1)^{k+r}}\!,
\end{equation}
where $\alpha=L\pi (r+1)/(L+2)$, and where we disentangle the indexes in Eq.~\eqref{GSM}, writing the general element as $\ket{k,(-1)^k}\ket{k+r,(-1)^{k+r}}, k=1,\ldots,L, r=0,\ldots,L'-1$. The periodicity of the indices with respect to $L'$ is implicit. 

To fix the normalization constant let us first introduce the following quantity
\begin{equation}\label{f_def}
	f_{d,b}^{m,n}(\alpha)=\sum_{r=m}^{n}\sin[\alpha(r+d+1)]\sin[\alpha(r+b+1)],
\end{equation}
with $\alpha \in \mathbb{R}$, and we write $f_{b,b}^{m,n}$ instead of $f_{b,b}^{m,n}(\alpha)$. Eq.~\eqref{f_def} admits a nice representation in terms of the Chebyshev polynomials of first and second type,  $T_n(\cos\alpha)$ and $U_n(\cos\alpha)$ respectively, remembering that $\sin[\alpha (n+1)] = \sin\alpha \; U_n(\cos\alpha)$ and resorting to the relations~\cite{Kilic2020} 
\begin{align}
	& T_{a+b}(x)-T_{a-b}(x)=2(x^2-1)U_{a-1}(x)U_{b-1}(x), \\
	& T_{2(a-b)}(x)-1=2(x^2-1)U^2_{a-b-1}(x), \\
	& T_{a+b}(x)+T_{a-b}(x) = 2T_a(x)T_b(x), \\
	& T_{2(a-b)(x)}+1 = 2T^2_{a-b}(x).
\end{align}
We have
\begin{align}\label{f_pol}
	f_{d,b}^{m,n} & =\frac{1}{2} (n-m+1) T_{b-d} + \nonumber \\
	& -\dfrac{1}{4} \left(U_{-(2m+b+d+2)}+ U_{2n+b+d+2}\right).
\end{align}
Furthermore we also need the square of Eq.~\eqref{f_def}
\begin{align}\label{f_pol_2}
	(f_{d,b}^{m,n})^2 & = \dfrac{(n-m+1)^2}{8}\left(1+T_{2(b-d)}\right) + \nonumber \\
	& + \dfrac{1}{32\sin^2(\alpha)}\left(2-T_{2(2n+b+d+3)}-T_{2(2m+b+d+1)}+ \right. \nonumber \\
	& \left. T_{2(n+m+b+d+2)-T_{2(n-m+1)}}\right) + \nonumber \\
	& - \dfrac{n-m+1}{16}\left(U_{2(n+b+1)}+U_{2(2n+d+1)}+ \right.\nonumber \\
	& \left. +U_{-2(n+d+1)}+U_{-2(n+b+1)}\right),
\end{align}
and we evaluate the asymptotic expression
\begin{equation}\label{T_lim}
	\lim_{\substack{L,M\to \infty \\ M/L \to C}}T_{a L + b M + c}(\cos\alpha) = (-1)^{b+c}\cos(2 b\pi x),
\end{equation}
with $x=M/L$, $\alpha = L \pi / (L+2)$, $0 < C < 1$, and $a,b,c \in \mathbb{Z}$, which is valid for even and odd $L, M \in \mathbb{N}$ respectively. 

Simplifying the notation as $T_n(\cos\alpha)=T_n$ and $U_n(\cos\alpha)=U_n$, the normalization constant can be then be calculated resulting in
\begin{equation}\label{eq:state_normalization}
	\braket{g}= \vert A\vert^2 L \, f_{0,0}^{0,\frac{L}{2}-1} = \vert A\vert^2 \left(\dfrac{L^2}{4}+\dfrac{L}{4}-\dfrac{L}{4}U_L\right)=1.
\end{equation}
Furthermore,  $U_L=-1$ for $L$ even. In fact, remembering that $\sin[\alpha (n+1)] = \sin\alpha \; U_n(\cos\alpha)$ we can prove equivalently that $    \sin\left[\alpha(L+1)\right] = - \sin\left(\alpha\right), \forall \;L$ even, condition that it is always satisfied for $L=4n+2, \forall n\in \mathbb{N}$. From Eq.~\eqref{eq:state_normalization} it then follows that \mbox{$A=2/\sqrt{L\left(L+2\right)}$}. The translational invariant form Eq.~\eqref{GS} is then obtained through the substitution $p\rightarrow k+r$.

\subsection{Asymptotic expression of the bipartite Entanglement Entropy close to the classical line.}\label{Asym_EE}

\begin{figure}
    \includegraphics[width=0.9\linewidth]{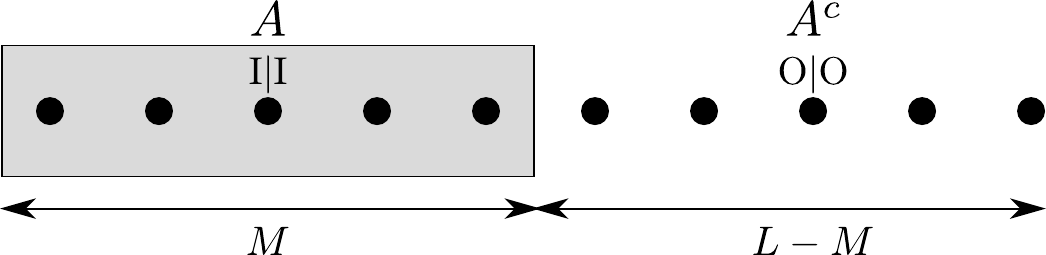}
    \caption{Partitioning of the ANNNI chain for the computation of the Entanglement Entropy Eq.~\eqref{eq:EE} near the classical point. The $\textrm{I}$ ($\textrm{O}$) notation indicates that the $(k, p):=k|p$ indexes are inside (outside) the partition $A$. }
    \label{fig:subpartition_plot}
\end{figure}

Having determined the expression for the ground state close to the classical line, in this appendix we detail the analytic derivation of its bipartite Entanglement Entropy. Without loos of generality, we consider a system sub-partition $A$ made by an odd number $M$ of consecutive spins. The corresponding spin number inside $A$ for the two sub-chains are then equal to $M_o=(M+1)/2$ and $M_e=(M-1)/2$ respectively (see Fig.~\ref{fig:subpartition_plot}).

As discussed in Sec.~\ref{pert_theory}, in analogy with the approaches used for the TF Ising chain~\cite{Torre2023},  it is useful to rewrite the state in Eq.~\eqref{GS} according to the kink indexes $(k, p):=k|p$ being internal $(\textrm{I})$ or external $(\textrm{O})$ to the subsystem $A$
\begin{equation}
    \ket{g} = \ket{\textrm{I}|\textrm{I}}+\ket{\textrm{I}|\textrm{O}}+\ket{\textrm{O}|\textrm{I}}+\ket{\textrm{O}|\textrm{O}},
\end{equation}
where the vertical bar $(\textrm{|})$ separates the indexes corresponding to the odd/even chains.  We can then express the corresponding reduced density matrix as a $16\times 16$ block matrix. Furthermore, we numerically observe that only four blocks contribute to the spectrum in the thermodynamic limit as shown in Fig.~\ref{fig:main_density_matrix_approximation}.
We can then consider the following block-diagonal approximated form for the reduced density matrix
\begin{equation}\label{approx_reduced_density_matrix}
    \rho_A = \left( 		\begin{array}{cccc}
			  \rho^{\textrm{I}\textrm{I}|\textrm{I}\textrm{I}} & 0 & 0 & 0 \\
			0 & \rho^{\textrm{I}\textrm{O}|\textrm{I}\textrm{O}} & 0 & 0 \\
			0 & 0 & \rho^{\textrm{O}\textrm{I}|\textrm{O}\textrm{I}} & 0 \\
			0 & 0 & 0 & \rho^{\textrm{O}\textrm{O}|\textrm{O}\textrm{O}} 
		\end{array} \right).
\end{equation}

Let us now discuss each block matrix and derive their eigenvalues separately.

\subsubsection{Asymptotic eigenvalues of the matrices $\rho^{\textrm{I}\textrm{I}|\textrm{I}\textrm{I}}$ and $\rho^{\textrm{O}\textrm{O}|\textrm{O}\textrm{O}}$}

We start considering the matrix $\rho^{\textrm{O}\textrm{O}|\textrm{O}\textrm{O}}$
\begin{equation}
	\rho^{\textrm{O}\textrm{O}|\textrm{O}\textrm{O}} = \dfrac{4}{2L+L^2}\sum_{n=0}^{\frac{L-M-1}{2}-1}f_{0,0}^{0,n}\mathbb{1}_{4 \times 4},
\end{equation}
where we have used the property $c_r=c_{L/2-r}$. We have then one eigenvalue with multiplicity four
\begin{align}\label{lambda1}
	&\lambda_{1}=\dfrac{4}{2L+L^2}\sum_{n=0}^{\frac{L-M-3}{2}}f_{0,0}^{0,n}\left(\alpha\right)= \nonumber \\
	& = \dfrac{(L-M-1)(L-M+3)}{4L(L+2)}+ \nonumber \\
    & - \dfrac{1}{2L(L+2)\sin^2\alpha}\left(T_2-T_{L-M+1}\right).
\end{align}
Its asymptotic expression is straightforward resorting to the limit Eq.~\eqref{T_lim}
\begin{equation}
    \lambda_{1}^{\mathrm{Asym}} (x) = \dfrac{(1-x)^2}{4}- \dfrac{\sin^2\pi x}{4\pi^2}.
\end{equation}

The matrix $\rho^{\textrm{I}\textrm{I}|\textrm{I}\textrm{I}}$ is a $4\times 4$ diagonal block matrix with identical blocks once resorting to the property $c_r^2=c_{L/2-r}^2$
\begin{equation}
	\rho^{\textrm{I}\textrm{I}|\textrm{I}\textrm{I}} = \bigoplus_{k=1}^4 \tilde{\rho}^{\textrm{I}\textrm{I}|\textrm{I}\textrm{I}},
\end{equation}
with the $(M+1)/2\times (M+1)/2$ matrix $\tilde{\rho}_{\textrm{I}\textrm{I}|\textrm{I}\textrm{I}}$ given by
\begin{equation}
	\tilde{\rho}^{\textrm{I}\textrm{I}|\textrm{I}\textrm{I}} = \left[v_{\frac{M-1}{2}},v_{\frac{M-3}{2}},\ldots,v_{0}\right]\otimes\left[v_{\frac{M-1}{2}},v_{\frac{M-3}{2}},\ldots,v_{0}\right],
\end{equation}
with $v_i=(c_0,c_1,\ldots,c_{i-1})$. Despite its complexity this matrix has rank one. From the Cayley-Hamilton theorem then the only non-zero eigenvalue with multiplicity four is given by its trace~\cite{Gantmacher}
\begin{align}
	\lambda_{2} & = \mathrm{Tr}\left[\tilde{\rho}^{\textrm{I}\textrm{I}|\textrm{I}\textrm{I}}\right] = \dfrac{4}{2L+L^2}\sum_{n=0}^{\frac{L-M-1}{2}-1}f_{0,0}^{0,n} = \nonumber \\
    & = \dfrac{(M-1)(M+3)}{4L(L+2)} + \nonumber \\
    & -\dfrac{1}{2L(L+2)\sin^2\alpha}\left(T_2-T_{M+1}\right).
\end{align}
Its asymptotic limit is given by
\begin{equation}\label{lambda2}
    \lambda_{2}^{\mathrm{Asym}} (x) = \dfrac{x^2}{4}-\dfrac{\sin^2\pi x}{4\pi^2}.
\end{equation}

\subsubsection{Asymptotic eigenvalues of the matrices $\rho^{\textrm{I}\textrm{O}|\textrm{I}\textrm{O}}$ and $\rho^{\textrm{O}\textrm{I}|\textrm{O}\textrm{I}}$}

The matrices $\rho^{\textrm{I}\textrm{O}|\textrm{I}\textrm{O}}$ and $\rho^{\textrm{O}\textrm{I}|\textrm{O}\textrm{I}}$ are $2\times 2$ block matrices with identical blocks of size respectively $\frac{M+1}{2}\times\frac{M+1}{2}$ and $\frac{M-1}{2}\times\frac{M-1}{2}$
\begin{equation}
	\rho^{\textrm{I}\textrm{O}|\textrm{I}\textrm{O}} = \bigoplus_{k=1}^2 \tilde{\rho}^{\textrm{I}\textrm{O}|\textrm{I}\textrm{O}},\quad \rho^{\textrm{O}\textrm{I}|\textrm{O}\textrm{I}} = \bigoplus_{k=1}^2 \tilde{\rho}^{\textrm{O}\textrm{I}|\textrm{O}\textrm{I}},
\end{equation}
with
\begin{align}\label{rho_IOIO_OIOI}
	& \left(\tilde{\rho}^{\textrm{I}\textrm{O}|\textrm{I}\textrm{O}}\right)_{i,j} = f_{\frac{L-M-1}{2}-i,\frac{L-M-1}{2}-j}^{0,\frac{L-M-1}{2}}, \nonumber \\
 & \left(\tilde{\rho}^{\textrm{O}\textrm{I}|\textrm{O}\textrm{I}}\right)_{i,j} =  f_{i,j}^{1,\frac{L-M-1}{2}},
\end{align}
where we resort again to the property $c_r=c_{L/2-r}$.

The matrices Eqs.~\eqref{rho_IOIO_OIOI} has rank two. As a consequence, resorting to the Cayley-Hamilton theorem, we can compute the non-zero eigenvalues solving the following quadratic equation~\cite{Gantmacher}
\begin{equation}
	\lambda^2-\mathrm{Tr}A \, \lambda-\dfrac{1}{2}\left[\mathrm{Tr}A^2-\left(\mathrm{Tr}A\right)^2\right]=0,
\end{equation}
with $ A= \tilde{\rho}^{\textrm{I}\textrm{O}|\textrm{I}\textrm{O}/\textrm{O}\textrm{I}|\textrm{O}\textrm{I}}$.
A straightforward computation thought Eq.s~\eqref{f_pol} and Eq.~\eqref{f_pol_2} shows that
\begin{widetext}
\begin{align}
	\mathrm{Tr}\tilde{\rho}_{\textrm{I}\textrm{O}|\textrm{I}\textrm{O}} & = \dfrac{4}{2L+L^2} \sum_{r=0}^{\frac{M-1}{2}} f_{0,0}^{r,\frac{L-M-1}{2}} = \dfrac{(M+1)(L-M+1)}{2L(L+2)} + \dfrac{1}{2L(L+2)\sin^2\alpha}\left(1-T_{M+1}-T_{L-M+1}+T_{L+2}\right).
\end{align}
\begin{align}
	\mathrm{Tr}(\tilde{\rho}^{\textrm{I}\textrm{O}|\textrm{I}\textrm{O}})^2 & = \sum_{b,d=0}^{\frac{M-1}{2}} \left(f_{d,b}^{0,\frac{L-M-1}{2}}\right)^2 = \dfrac{(L-M+1)^2(M+1)^2}{8L^2(L+2)^2} + \dfrac{(L-M+1)^2}{4L^2(L+2)^2\sin^2\alpha}(1-T_{M+1}) + \nonumber \\
    & - \dfrac{(L-M+1)(M+1)}{2L^2(L+2)^2\sin^2\alpha}(T_{L-M+1}-T_{L+2}+T_{M+1}-1) + \dfrac{(M+1)^2}{4L^2(L+2)^2\sin^2\alpha}(1-T_{L-M+1}) + \nonumber \\
    & + \dfrac{1}{4L^2(L+2)^2\sin^4\alpha}\left[2T_{L+2}-T_{L-M+3}-T_{L-M+1}-T_{2L-M+3}-T_{M+1}+\dfrac{1}{2}(T_{2(L-M+1)}+T_{2(L+2)}+T_{2(M+1)})\right],
\end{align}
\begin{align}
	\mathrm{Tr}\tilde{\rho}^{\textrm{O}\textrm{I}|\textrm{O}\textrm{I}} & =\dfrac{4}{2L+L^2}\sum_{r=1}^{\frac{M-1}{2}} f_{0,0}^{r,r+\frac{L-M-3}{2}} = \dfrac{(M-1)(L-M-1)}{2L(L+2)} + \dfrac{1}{2L(L+2)\sin^2\alpha}\left[T_2-T_{M+1}-T_{L-M+1}+T_{L}\right],
\end{align}
\begin{align}
	\mathrm{Tr}(\tilde{\rho}^{\textrm{O}\textrm{I}|\textrm{O}\textrm{I}})^2 & = \sum_{b,d=0}^{\frac{M-3}{2}} \left(f_{d,b}^{1,\frac{L-M-1}{2}}\right)^2 = \dfrac{(L-M-1)^2(M-1)^2}{8L^2(L+2)^2} + \dfrac{(L-M-1)^2}{4L^2(L+2)^2\sin^2\alpha}(1-T_{L-M+1}) + \nonumber \\
    & - \dfrac{(L-M-1)(M-1)}{2L^2(L+2)^2\sin^2\alpha}(T_{L-M+1}-T_L+T_{M+1}-T_2) + \dfrac{(M-1)^2}{4L^2(L+2)^2\sin^2\alpha}(1-T_{L-M-1}) + \nonumber \\
    & + \dfrac{1}{4L^2(L+2)^2\sin^4\alpha}\left[2T_{L+2}-T_{L-M+3}-T_{L+M+1}-T_{2L-M+1}-T_{M+3}+\dfrac{1}{2}(T_{2(L-M+1)}+T_{2L}+T_{2(M+1)})+T_4\right].
\end{align} 
\end{widetext}
Through the limit Eq.~\eqref{T_lim} we obtain the same asymptotic expression for the eigenvalues of the two matrices, with multiplicity four
\begin{align}\label{lambda34}
    & \lambda_{3,4}^{\mathrm{Asym}}(x) = \dfrac{x(1-x)}{4} + \dfrac{\sin^2\pi x}{4\pi^2} \pm \dfrac{\sin\pi x}{4 \pi}.
\end{align}





\subsubsection{Collecting the reduced density matrix eigenvalues}

The asymptotic expression for the entanglement entropy Eq.~\eqref{EE_anal_result} can be obtained through the definition Eq.~\eqref{eq:EE} and the asymptotic expression of the eigenvalues Eq.~\eqref{eig_limit}, as well as with Eq.s~\eqref{lambda1},~\eqref{lambda2}, and~\eqref{lambda34}, noticing that
\begin{align}
	& \lambda_{1}^{\mathrm{Asym}} = \dfrac{1}{4} (1-y)(1-z), \quad \lambda_{2}^{\mathrm{Asym}} = \dfrac{1}{4} yz, \nonumber \\
	& \lambda_{3}^{\mathrm{Asym}} = \dfrac{1}{4} y(1-z), \quad \lambda_{4}^{\mathrm{Asym}} = \dfrac{1}{4} (1-y)z,
\end{align}
with $y=x-\frac{\sin\pi x}{\pi}$ and $z=x+\frac{\sin\pi x}{\pi}$.

\section{Numerical analysis}
\label{ap:Numerics}

\begin{figure}[t]
    \includegraphics[width=.99\columnwidth]{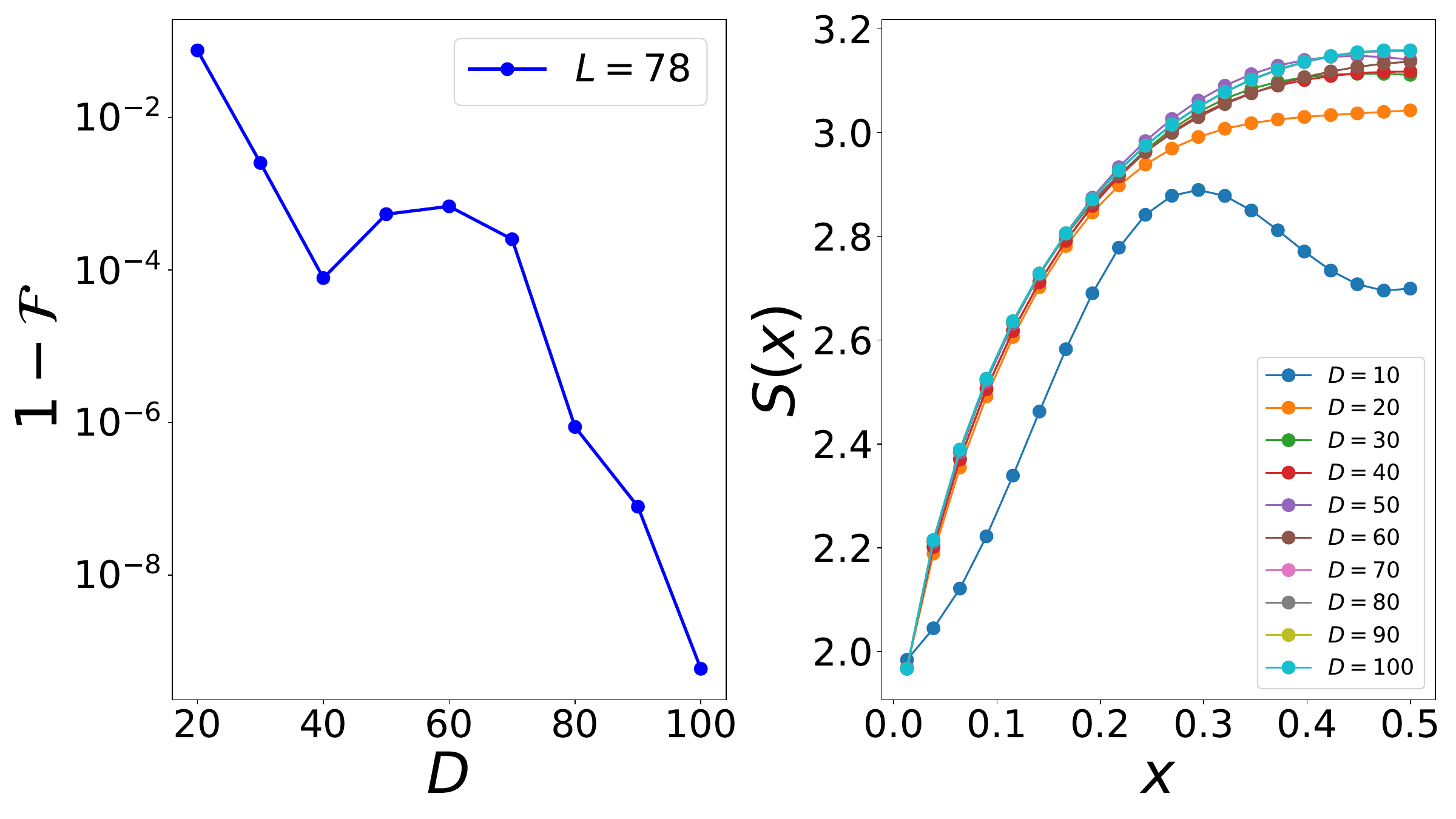}
    \caption{Left: scaling of the infidelity between ground-states computed with different bond dimension. Right: saturation of the entanglement entropy with increasing bond dimension $D$.}
    \label{fig:BondD}
\end{figure}

We performed a numerical estimation of the ground-state of the Hamiltonian Eq.~\eqref{eq:Hamiltonian} using the density matrix renormalization group (DMRG) technique. In particular, we implemented a DMRG based on tensor networks, following the approach of \cite{Catarina23}. Despite the fact that in presence of topological frustration the EE is expected to violate the area law, it is also known from \cite{Giampaolo2019} that it will saturate to a finite value for large system's sizes, as opposite to what happens in critical systems where a logarithmic divergence of the EE is expected in thermodynamic limit. Therefore, one could expect that an MPS ansatz can still reproduce the ground-states of topologically frustrated systems in an efficient way.

In order to benchmark the validity of the MPS ansatz we computed the fidelity between ground-states corresponding to MPS with different bond-dimensions $D$
\begin{equation}
\mathcal{F}=|\bra{GS(D+\Delta D)}\ket{GS(D)}|,
\end{equation}
and checked the saturation of the entanglement entropy as a function of the bond dimension.  
The results are shown in Fig.~\ref{fig:BondD}. We observe that the infidelity, i.e. $1-\mathcal{F}$, decays with increasing bond-dimension. Moreover, also the EE entropy shows a saturation for increasing bond-dimension, confirming the convergence of the DMRG algorithm. for sufficiently large bond-dimension.

Based on the results of Fig. \ref{fig:BondD} we decided to use $D=70$ to collect the data presented in Sec. \ref{num_analys}.


\begin{thebibliography}{99}

\bibitem{Landau1937} L. D. Landau. On the Theory of Phase Transitions, Zh. Eksp. Teor. Fiz. \textbf{7}: 19-32 (1937). doi: \href{https://doi.org/10.1016/B978-0-08-010586-4.50034-1}{10.1016/B978-0-08-010586-4.50034-1}

\bibitem{Wen&Niu1990}
X.-G. Wen, and Q. Niu, Ground state degeneracy of the FQH states in presence of random potential and on high genus Riemann surfaces, Phys. Rev. B. \textbf{41}, 9377 (1990). doi: \href{https://doi.org/10.1103/PhysRevB.41.9377}{10.1016/B978-0-08-010586-4.50034-1}

\bibitem{Wen1990}
X.-G. Wen, Topological Orders in Rigid States, Int. J. Mod. Phys. B. \textbf{4}, 239 (1990). doi: \href{https://doi.org/10.1142/S0217979290000139}{10.1142/S0217979290000139}




\bibitem{Zeng2015} B. Zeng, X. Chen, D.-L. Zhou,  and  X.-G.  Wen. Quantum Information Meets Quantum Matter: From Quantum Entanglement to Topological Phases of Many-Body Systems, (Springer, 2019). doi: \href{https://doi.org/10.1007/978-1-4939-9084-9}{10.1007/978-1-4939-9084-9}

\bibitem{Stanescu2016} T. d. Stanescu, Introduction to Topological Quantum Matter and Quantum Computation, (CRC Press, 2016). doi: \href{https://doi.org/10.1201/9781315181509}{10.1201/9781315181509}

\bibitem{Moessner2021} R. Moessner, and J.E. Moore, Topological Phases of Matter, (Cambridge University Press, 2021). doi: \href{    https://doi.org/10.1017/9781316226308}{10.1017/9781316226308}

\bibitem{Wen2002} X.-G. Wen, Quantum order: a quantum entanglement of many particles, Physics Letters A \textbf{300}, 175–181 (2002). doi: \href{https://doi.org/10.1016/S0375-9601(02)00808-3}{10.1016/S0375-9601(02)00808-3}

\bibitem{Chen2010} X. Chen, Z.-C. Gu, and X.-G. Wen, Local unitary transformation, long-range quantum entanglement, wave function renormalization, and topological order, Phys. Rev. B \textbf{82}, 155138 (2010). doi: \href{https://doi.org/10.1103/PhysRevB.82.155138}{10.1103/PhysRevB.82.155138}

\bibitem{Wen2013} X.-G. Wen, Topological Order: From Long-Range Entangled Quantum Matter to a Unified Origin of Light and Electrons, International Scholarly Research Notices 2013, e198710 (2013). doi: \href{https://doi.org/10.1155/2013/198710}{10.1155/2013/198710}

\bibitem{Wen2017} X.-G. Wen, Colloquium: Zoo of quantum-topological phases of matter, Rev. Mod. Phys. \textbf{89}, 041004 (2017). doi: \href{https://doi.org/10.1103/RevModPhys.89.041004}{10.1103/RevModPhys.89.041004}

\bibitem{Wen2019} X.-G. Wen, Choreographed entanglement dances: Topological states of quantum matter, Science \textbf{363}, eaal3099 (2019). doi: \href{https://doi.org/10.1126/science.aal3099}{10.1126/science.aal3099}

\bibitem{Hamma2005} A. Hamma, R. Ionicioiu, and P. Zanardi, Bipartite entanglement and entropic boundary law in lattice spin systems, Phys. Rev. A \textbf{71}, 022315 (2005). doi: \href{https://doi.org/10.1103/PhysRevA.71.022315}{/10.1103/PhysRevA.71.022315}

\bibitem{Shor2016}
R. Movassagh, and P. W. Shor, Supercritical entanglement in local systems: counterexample to the area law for quantum matter, Proc. Natl Acad. Sci. USA \textbf{113}, 13278 (2016). doi: \href{https://doi.org/10.1073/pnas.1605716113}{10.1073/pnas.1605716113}


\bibitem{Tong2021}
B. Tong, O. Salberger, K. Hao, and V. Korepin, Shor–Movassagh chain leads to unusual integrable model, Journal of Physics A: Mathematical and Theoretical \textbf{54} 394002 (2021). doi: \href{https://doi.org/10.1088/1751-8121/ac1f3f}{10.1088/1751-8121/ac1f3f}

\bibitem{fact1}
S. M. Giampaolo, G. Adesso, and F. Illuminati, Theory of ground state factorization in quantum cooperative systems,
Physical Review Letters {\bf 100}, 197201 (2008).doi: \href{https://doi.org/10.1103/PhysRevLett.100.197201}{10.1103/PhysRevLett.100.197201}

\bibitem{fact2}
S. M. Giampaolo, G. Adesso, and F. Illuminati, Separability and ground-state factorization in quantum spin systems, Physical Review B {\bf 79}, 224434 (2009). doi \href{https://doi.org/10.1103/PhysRevB.79.224434}{10.1103/PhysRevB.79.224434}

\bibitem{fact3}
S. M. Giampaolo, G. Adesso, and F. Illuminati, Probing Quantum Frustrated Systems via Factorization of the Ground State,  
Physical Review Letters {\bf 104}, 207202 (2010). doi:\href{https://doi.org/10.1103/PhysRevLett.104.207202}{10.1103/PhysRevLett.104.207202}

\bibitem{Wolf2003} M. M. Wolf, F. Verstraete, J. I. Cirac, { Entanglement and Frustration in Ordered Systems}, Int. Journal of Quantum Information 1, 465 (2003). doi: \href{https://doi.org/10.1142/S021974990300036X}{10.1142/S021974990300036X}

\bibitem{Giampaolo2011} S. M. Giampaolo, G. Gualdi, A. Monras, F. Illuminati, { Characterizing and quantifying frustration in quantum many-body systems}, Phys. Rev. Lett. \textbf{107}, 260602 (2011). doi: \href{https://doi.org/10.1103/PhysRevLett.107.260602}{10.1103/PhysRevLett.107.260602}

\bibitem{Giampaolo2013} U. Marzolino, S. M. Giampaolo, F. Illuminati, { Frustration, entanglement, and correlations in quantum many body systems}, Phys. Rev. A \textbf{88}, 020301(R) (2013). doi: \href{https://doi.org/10.1103/PhysRevA.88.020301}{10.1103/PhysRevA.88.020301}

\bibitem{Giampaolo2015}
S. M. Giampaolo, B. C. Hiesmayr, and F. Illuminati, Global-to-local incompatibility, monogamy of entanglement, and ground-state dimerization: Theory and observability of quantum frustration in systems with competing interactions, Phys. Rev. B \textbf{92}, 144406 (2015). doi: \href{https://doi.org/10.1103/PhysRevB.92.144406}{10.1103/PhysRevB.92.144406}

\bibitem{Toulouse1977} G. Toulouse, { Theory of the frustration effect in spin glasses: I}, Commun. Phys. \textbf{2}, 115 (1977). doi: \href{https://doi.org/10.1142/9789812799371_0009}{10.1142/9789812799371\_0009}

\bibitem{Vannimenus1977} J. Vannimenus, G. Toulouse, { Theory of the frustration effect. II. Ising spins on a square lattice},
J. Phys. C \textbf{10}, L537 (1977). doi: \href{https://doi.org/10.1088/0022-3719/10/18/008}{10.1088/0022-3719/10/18/008}

\bibitem{Dong2016}
J.-J. Dong, P. Li, \and Q. -H. Chen, The A-Cycle Problem for Transverse Ising Ring, J. Stat. Mech. P113102 (2016). doi: \href{https://doi.org/10.1088/1742-5468/2016/11/113102}{10.1088/1742-5468/2016/11/113102}

\bibitem{Sen2008}
A. Sen, U. Sen, J. Dziarmaga, A. Sanpera, and M. Lewenstein, { Frustration, Area Law, and Interference in Quantum Spin Models}, Phys. Rev. Lett. \textbf{101}, 187202 (2008). doi: \href{https://doi.org/10.1103/PhysRevLett.101.187202}{10.1103/PhysRevLett.101.187202}

\bibitem{Maric2022_fate} V. Mari{\'c}, S. M. Giampaolo, and Fabio Franchini, { Fate of local order in topologically frustrated spin chains}, Physical Review B {\bf 105}, 064408 (2022). doi: \href{https://doi.org/10.1103/PhysRevB.105.064408}{10.1103/PhysRevB.105.064408}

\bibitem{Catalano2022} A. G. Catalano, D. Brtan, F. Franchini and S. M. Giampaolo, { Simulating continuous symmetry models with discrete ones}, Phys. Rev. B, \textbf{106}, 125145 (2022). doi: \href{https://doi.org/10.1103/PhysRevB.106.125145}{10.1103/PhysRevB.106.125145}

\bibitem{Odavic2023}
J. Odavi\'c, T. Haug, G. Torre, A. Hamma, F. Franchini, and S. M. Giampaolo, { Complexity of frustration: a new source of non-local non-stabilizerness}, SciPost Phys. \textbf{15}, 131 (2023). doi: \href{https://doi.org/10.21468/SciPostPhys.15.4.131}{10.21468/SciPostPhys.15.4.131}

\bibitem{Maric2020_destroy} V. Mari{\'c}, S. M. Giampaolo and F. Franchini, { The frustration of being odd: how boundary conditions can destroy local order}, New Journal of Physics, \textbf{22}, 083024 (2020). doi: \href{https://doi.org/10.1088/1367-2630/aba064}{10.1088/1367-2630/aba064}

\bibitem{Maric2020_neworder}  V. Mari{\'c}, S. M. Giampaolo and F. Franchini, 
{ Quantum Phase Transition induced by topological frustration}, Communications Physics {\bf 3}, 220 (2020). doi: \href{https://doi.org/10.1038/s42005-020-00486-z}{10.1038/s42005-020-00486-z}

\bibitem{Torre2022} G. Torre, V. Mari\'c, D. Kui\'c, F. Franchini, and S. M. Giampaolo, { Odd thermodynamic limit for the Loschmidt echo}, Physical Review B, \textbf{105}, 184424 (2022). doi: \href{https://doi.org/10.1103/PhysRevB.105.184424}{10.1103/PhysRevB.105.184424}

\bibitem{Giampaolo2019} S. M. Giampaolo, F. B. Ramos and F. Franchini, { The Frustration of being Odd: Universal area law violation in local systems}, Journal of Physics Communication {\bf 3}, 081001 (2019). doi: \href{https://doi.org/10.1088/2399-6528/ab3ab3}{10.1088/2399-6528/ab3ab3}


\bibitem{Torre2023}
G. Torre, J. Odavi\'c, P. Fromholz, S. M. Giampaolo, and F. Franchini, Long-range entanglement and topological excitations, 
arXiv:2310.16091 (2023). doi: \href{https://arxiv.org/abs/2310.16091}{arXiv:2310.16091}







\bibitem{Selke88} W. Selke, { The ANNNI model - theoretical analysis and experimental application}, Sci. Rep.,  \textbf{170}, 4 (1988).




\bibitem{Suzuki2013} S. Suzuki, J. Inoue, B. K. Chakrabarti, { Quantum Ising Phases and Transitions in Transverse Ising Models}, Lecture Notes in Physics \textbf{862}, Springer (2013). doi: \href{https://doi.org/10.1007/978-3-642-33039-1}{10.1007/978-3-642-33039-1}


\bibitem{Monaco2023}
S. Monaco, O. Kiss, A. Mandarino, S. Vallecorsa, and M. Grossi, Phys. Rev. B \textbf{107}, L081105 (2023). doi: \href{https://doi.org/10.1103/PhysRevB.100.045129}{10.1103/PhysRevB.100.045129} 

\bibitem{Canabarro2019} A. Canabarro, F. F. Fanchini, A. L. Malvezzi, R. Pereira, R. Chaves, { Unveiling phase transitions with machine learning}, Phys. Rev. B \textbf{100}, 045129 (2019). doi: \href{https://doi.org/10.1103/PhysRevB.100.045129}{10.1103/PhysRevB.100.045129}

\bibitem{Zippilli2014}
S. Zippilli, M. Johanning, S. M. Giampaolo, Ch. Wunderlich, and F. Illuminati, Adiabatic quantum simulation with a segmented ion trap: Application to long-distance entanglement in quantum spin systems, Phys. Rev. A \textbf{89}, 042308 (2014). doi: \href{https://doi.org/10.1103/PhysRevA.89.042308}{10.1103/PhysRevA.89.042308}


\bibitem{Gantmacher}
F. R. Gantmacher, \emph{The theory of matrices, vol. 1}, AMS Chelsea Publishing, 2000.


\bibitem{Beccaria06} M. Beccaria, M. Campostrini, A. Feo, { Density-matrix renormalization-group study of the disorder line in the quantum axial next-nearest-neighbor Ising model}, Phys. Rev. B \textbf{73}, 052402 (2006). doi: \href{https://doi.org/10.1103/PhysRevB.73.052402}{10.1103/PhysRevB.73.052402}

\bibitem{Beccaria07} M. Beccaria, M. Campostrini, A. Feo, { Evidence for a floating phase of the transverse ANNNI model at high frustration}, Phys. Rev. B, \textbf{76}, 094410 (2007). doi: \href{https://doi.org/10.1103/PhysRevB.76.094410}{10.1103/PhysRevB.76.094410} 

\bibitem{Bak1982} P. Bak, { Commensurate phases, incommensurate phases and the devil's staircase}, Rep. Prog. Phys. 45, 587 (1982). doi: \href{https://doi.org/10.1088/0034-4885/45/6/001}{10.1088/0034-4885/45/6/001}

\bibitem{Giamarchi} T. Giamarchi, { Quantum physics in one dimension}, Oxford University Pres (2003). doi: \href{https://doi.org/10.1093/acprof:oso/9780198525004.001.0001}{10.1093/acprof:oso/9780198525004.001.0001}

\bibitem{VN1} I. Bengtsson, and K. Zyczkowski, Geometry of Quantum States: An Introduction to Quantum Entanglement, Cambridge University Press (2006). doi: \href{    https://doi.org/10.1017/CBO9780511535048}{10.1017/CBO9780511535048}

\bibitem{VN2} M. A. Nielsen, and I. Chuang,  Quantum computation and quantum information, Cambridge University Press (2001). doi: \href{ https://doi.org/10.1017/CBO9780511976667 }{10.1017/CBO9780511976667 }


\bibitem{Berkovits2013}
R. Berkovits, Two particle excited states entanglement entropy in a one-dimensional ring, Phys. Rev. B \textbf{87}, 075141 (2013). doi: \href{https://doi.org/10.1103/PhysRevB.87.075141}{10.1103/PhysRevB.87.075141}



\bibitem{White92} S. R. White, { Density matrix formulation for quantum renormalization groups}, Phys. Rev. Lett. \textbf{69}, 2863 (1992). doi: \href{https://doi.org/10.1103/PhysRevLett.69.2863}{10.1103/PhysRevLett.69.2863}

\bibitem{Catarina23} G. Catarina, B. Murta, Density-matrix renormalization group: a pedagogical introduction., Eur. Phys. J. B \textbf{96}, 111 (2023). doi: \href{https://doi.org/10.1140/epjb/s10051-023-00575-2}{10.1140/epjb/s10051-023-00575-2}

\bibitem{Orus14} R. Or\'us, { A Practical Introduction to Tensor Networks: Matrix Product States and Projected Entangled Pair States}, Annals of Physics \textbf{349}, 117-158 (2014)

\bibitem{Orus19} R. Or\'us, { Tensor networks for complex quantum systems}, Nature Reviews Physics \textbf{1}, 538–550 (2019). doi: \href{https://doi.org/10.1038/s42254-019-0086-7}{10.1038/s42254-019-0086-7}

\bibitem{Biamonte20} J. Biamonte, { Lectures on Quantum Tensor Networks}, arXiv:1912.10049v2 (2020). doi: \href{https://arxiv.org/abs/1912.10049}{arXiv:1912.10049}

\bibitem{Weyrauch2013} M. Weyrauch, M. V. Rakov, { Efficient MPS algorithm for periodic boundary conditions and applications}, Ukr. J. Phys., Vol. 58, No. 7, pages 657-665 (2013). doi: \href{https://doi.org/10.15407/ujpe58.07.0657}{10.15407/ujpe58.07.0657}

\bibitem{Verstraete2004} F. Verstraete, D. Porras, and J. I. Cirac, { Density Matrix Renormalization Group and Periodic Boundary Conditions: A Quantum Information Perspective}, Phys. Rev. Lett. \textbf{93}, 227205 (2004). doi: \href{https://doi.org/10.1103/PhysRevLett.93.227205}{10.1103/PhysRevLett.93.227205}

\bibitem{Pippan2010} P. Pippan, S. R. White, and H. G. Evertz, { Efficient matrix-product state method for periodic boundary conditions}, Phys. Rev. B \textbf{81}, 081103(R) (2010). doi: \href{https://doi.org/10.1103/PhysRevB.81.081103}{10.1103/PhysRevB.81.081103}

\bibitem{Vidal2003} G. Vidal, J.I. Latorre, E. Rico, and A. Kitaev, Entanglement in Quantum Critical Phenomena, Phys. Rev. Lett. \textbf{90}, 227902 (2003). doi: \href{https://doi.org/10.1103/PhysRevLett.90.227902}{10.1103/PhysRevLett.90.227902}

\bibitem{Latorre2004} J.I. Latorre, E. Rico, and G. Vidal, Ground state entanglement in quantum spin chains, Quant. Inf. Comp. \textbf{4}, 48 (2004). doi: \href{https://doi.org/10.26421/QIC4.1-4}{10.26421/QIC4.1-4}

\bibitem{Eisert2010} J. Eisert, M. Cramer, and M. B. Plenio, Area laws for the entanglement entropy - a review, Rev. Mod. Phys. \textbf{82}, 277 (2010). doi: \href{https://doi.org/10.1103/RevModPhys.82.277}{10.1103/RevModPhys.82.277}


\bibitem{lee93}
    S.L. Lee and Y.N. Yeh,
    On Eigenvalues and Eigenvectors of Graphs,
    Journal of Mathematical Chemistry, \textbf{12}, 121-135 (1993).
    doi: \href{https://doi.org/10.1007/BF01164630}{10.1007/BF01164630}

\bibitem{Clancy20}
K. Clancy, M. Haythorpe, and A. Newcombe, A survey of graphs with known or bounded crossing numbers, Australasian journal of combinatorics \textbf{78(2)}, 209-296 (2020).

\bibitem{Barik18} S. Barik, D. Kalita, S. Pati, and G. Sahoo,
    Spectra of Graphs Resulting from Various Graph Operations and Products: A Survey,
    Special Matrices \textbf{6}, 342 (2018).
    doi: \href{https://doi.org/10.1515/spma-2018-0027}{10.1515/spma-2018-0027}


\bibitem{Kilic2020} Emrah K\i l\i \c{c}, Sibel Koparal, Ne\c{s}e \"{O}m\"{u}r, Powers Sums of the First and Second Kinds of Chebyshev Polynomials, Iran. J. Sci. Technol. Trans. A: Sci., \textbf{44}, 2 (2020). doi: \href{https://doi.org/10.1007/s40995-020-00843-1}{10.1007/s40995-020-00843-1}




\end{thebibliography}
\end{document}